\documentclass[useAMS,a4,usenatbib]{mn2e}
\input psfig.sty

\def \xoff {\ifmmode x_{\rm off} \else $x_{\rm off}$ \fi}
\def \rhorms {\ifmmode \rho_{\rm rms} \else $\rho_{\rm rms}$ \fi}

\def \apj  {ApJ}
\def \apjs  {ApJS}
\def \prd {Phy.Rev.D}
\def \mnras {MNRAS}
\def \etal {et~al.~}
\def \chisq  {\ifmmode  \chi^2   \else  $\chi^2$  \fi}  
\def \spose#1{\hbox  to 0pt{#1\hss}}  
\def \lta{\mathrel{\spose{\lower 3pt\hbox{$\sim$}}\raise  2.0pt\hbox{$<$}}}
\def \gta{\mathrel{\spose{\lower  3pt\hbox{$\sim$}}\raise 2.0pt\hbox{$>$}}}

\def \kms {\ifmmode  \,\rm km\,s^{-1} \else $\,\rm km\,s^{-1}  $ \fi }
\def \kpc {\ifmmode  {\rm kpc}  \else ${\rm  kpc}$ \fi  }  
\def \Msun {\ifmmode M_{\odot} \else $M_{\odot}$ \fi} 
\def \hMsun {\ifmmode h^{-1}\,\rm M_{\odot} \else $h^{-1}\,\rm M_{\odot}$ \fi}

\def \LCDM {\ifmmode \Lambda{\rm CDM} \else $\Lambda{\rm CDM}$ \fi}
\def \sig8 {\ifmmode \sigma_8 \else $\sigma_8$ \fi} 
\def \OmegaM {\ifmmode \Omega_{\rm M} \else $\Omega_{\rm M}$ \fi} 
\def \OmegaL {\ifmmode \Omega_{\rm \Lambda} \else $\Omega_{\rm \Lambda}$\fi} 
\def \Deltavir {\ifmmode \Delta_{\rm vir} \else $\Delta_{\rm vir}$ \fi}
\def \rhocrit {\ifmmode \rho_{\rm crit} \else $\rho_{\rm crit}$ \fi}
\def \rhou {\ifmmode \rho_{\rm u} \else $\rho_{\rm u}$ \fi}
\def \zc {\ifmmode z_{\rm c} \else $z_{\rm c}$ \fi}

\def \rhos {\ifmmode \rho_{\rm s} \else $\rho_{\rm s}$ \fi} 
\def \rs {\ifmmode r_{\rm s} \else $r_{\rm s}$ \fi} 
\def \cvir {\ifmmode c_{\rm vir} \else $c_{\rm vir}$ \fi} 
\def \Rvir {\ifmmode r_{\rm vir} \else $R_{\rm vir}$ \fi}
\def \Vvir {\ifmmode V_{\rm  vir} \else  $V_{\rm vir}$  \fi} 
\def \Mvir {\ifmmode M_{\rm  vir} \else $M_{\rm  vir}$ \fi}  
\def \Nvir {\ifmmode N_{\rm  vir} \else $N_{\rm  vir}$ \fi}  
\def \Jvir {\ifmmode J_{\rm vir} \else $J_{\rm vir}$ \fi} 
\def \Evir {\ifmmode E_{\rm vir} \else $E_{\rm vir}$ \fi} 
\def \lam {\ifmmode \lambda  \else $\lambda$ \fi} 
\def \lamp {\ifmmode \lambda^{\prime} \else $\lambda^{\prime}$  \fi} 
\def \Vmax {\ifmmode V_{\rm  max} \else  $V_{\rm max}$  \fi} 

\title[Cosmological model and  Dark Matter halos] {Concentration, Spin
  and Shape  of Dark Matter Haloes  as a Function  of the Cosmological
  Model: WMAP1, WMAP3 \& WMAP5 results.}

\author[A.V.    Macci\`o   et   al.]    {Andrea   V.    Macci\`o$^{1}$
  \thanks{maccio@mpia.de}, Aaron A.  Dutton$^2$, Frank C. van den 
  Bosch$^1$\\ 
  $^1$Max-Planck-Institut f\"ur Astronomie, K\"onigstuhl 17, 69117 
  Heidelberg, Germany \\ 
  $^2$UCO/Lick   Observatory   and   Department   of   Astronomy   and 
  Astrophysics, University of California, Santa Cruz, CA \\ 
}

\begin{document} 
              
\date{} 
              
\pagerange{\pageref{firstpage}--\pageref{lastpage}}\pubyear{2008} 
 
\maketitle            

\label{firstpage}
             
\begin{abstract}
  
  We investigate the effects of changes in the cosmological parameters
  between the WMAP 1st, 3rd, and  5th year results on the structure of
  dark  matter haloes.   We  use a  set  of simulations  that cover  5
  decades in halo mass ranging from the scales of dwarf galaxies ($V_c
  \approx 30$ km/s) to clusters of galaxies ($V_c \approx 1000$ km/s).
  We find that  the concentration mass relation is a  power law in all
  three  cosmologies.  However  the slope  is shallower  and  the zero
  point is lower  moving from WMAP1 to WMAP5 to  WMAP3.  For haloes of
  mass  $\log   M_{200}/[\hMsun]$  =  10,  12,   and  14  the
  differences in  the concentration parameter between  WMAP1 and WMAP3
  are a factor of 1.55, 1.41, and 1.29, respectively. As we show, this
  brings the central densities of dark matter haloes in good agreement
  with  the central  densities  of dwarf  and  low surface  brightness
  galaxies inferred from their rotation curves, for both the WMAP3 and
  WMAP5 cosmologies. We also show that none of the existing toy models
  for  the concentration-mass  relation can  reproduce  our simulation
  results over the entire range  of masses probed.  In particular, the
  model of  Bullock \etal (2001a;  hereafter B01) fails at  the higher
  mass end $(M \gta 10^{13}  \hMsun$), while the NFW model of Navarro,
  Frenk \&  White (1997)  fails dramatically at  the low mass  end ($M
  \lta 10^{12}  \hMsun$).  We present a  new model, based  on a simple
  modification of that of B01, which reproduces the concentration-mass
  relations in our simulations over  the entire range of masses probed
  ($10^{10} \hMsun \lta M \lta 10^{15} \hMsun$).   
  Haloes in the WMAP3 cosmology (at a fixed mass) are more flatted 
  compared to the WMAP1 cosmology, with a medium to long axis ration 
  reduced by $\approx 10\%$. 
  Finally, we show that the distribution of  halo spin parameters is 
  the  same for all three cosmologies.

\end{abstract}


\setcounter{footnote}{1}

\section{Introduction}
\label{sec:intro}

In the paradigm of  hierarchical structure formation, dark matter (DM)
haloes provide the potential well in which galaxies subsequently form.
As a consequence the structural  parameters of disk galaxies (size and
rotation velocity) are tightly coupled  with those of their hosting DM
halo,  such as concentration  and spin  (e.g. Mo,  Mao \&  White 1998;
Dutton \etal 2007).

It has been shown by several studies that the structural properties of
dark matter haloes are dependent on halo mass: for example higher mass
halos are  less concentrated (Navarro, Frank \&  White 1997, hereafter
NFW;  Bullock \etal  2001a, hereafter  B01; Eke  Navarro  \& Steinmetz
2001;  Kuhlen  \etal 2005;  Neto  \etal  2007;  Macci\`o \etal  2007,
hereafter M07), and are more prolate (Jing \& Suto 2002; Allgood \etal
2006; M07)  on average.  In the case  of the spin  parameter, however,
there seems to be no mass dependence.

In M07 we  used a set of numerical  simulations of structure formation
in a  $\Lambda$CDM cosmology,  with cosmological parameters  that were
motivated  by  the  first  year  results of  the  Wilkinson  Microwave
Anisotropy Probe (WMAP) mission (Spergel \etal 2003; hereafter WMAP1),
to study how  concentrations, shapes  and spin  parameters of
dark matter haloes  scale with halo mass. In this  paper we extent the
M07 study by  investigating how these scaling relations  depend on the
adopted  cosmology.   In  particular,  we  present a  large  suite  of
numerical  simulations for  $\Lambda$CDM cosmologies  whose parameters
are motivated  by the three  and five year  data releases of  the WMAP
mission (Spergel \etal  2007; Komatsu \etal 2008). In  what follows we
refer  to  these  cosmologies  as  the WMAP3  and  WMAP5  cosmologies,
respectively. The parameters of the WMAP1, WMAP3 and WMAP5 cosmologies
are listed in  Table \ref{tab:params}.  Note that the  WMAP3 and WMAP5
cosmologies both have a lower  matter density, $\Omega_m$, and a lower
power spectrum  normalization, $\sigma_8$ (defined  as the rms  of the
matter field, linearly  extrapolated to the present, on  a scale of $8
h^{-1}  \rm Mpc$)  than the  WMAP1 cosmology.  This implies  that dark
matter  haloes of a  given mass  assemble later  (e.g., van  den Bosch
2001).  Since  the concentration of a  dark matter halo  is related to
its assembly redshift (e.g.,  Wechsler \etal 2002; Zhao \etal 2003a,b;
Li  \etal  2007),   dark  matter  haloes  are  expected   to  be  less
concentrated in  the WMAP3  and WMAP5 cosmologies.   This may  help to
reconcile the  conflict between the  concentration parameters inferred
from the  rotation curves  of dwarf and  low surface  brightness (LSB)
galaxies  (de Blok,  McGaugh \&  Rubin 2001;  Swaters \etal  2003) and
those  predicted  from  the  \LCDM  scenario.  In  addition,  a  lower
amplitude of the  power spectrum results in haloes  being more prolate
(Allgood \etal  2006), which  may have observational  implications for
the rotation curves of dark matter dominated galaxies.

As in  M07 we use a large  suite of N-body simulations  for the WMAP1,
WMAP3  and WMAP5  cosmologies with  different box  sizes to  cover the
entire halo mass  range from $10^{10} \hMsun$ (haloes  that host dwarf
galaxies)  to $10^{15}  \hMsun  $ (massive  clusters).   We use  these
simulations  to  investigate the  concentrations,  spin parameters  of
shapes of dark matter haloes. We also present a critical comparison of
the simulation results with  two toy models for the concentration-mass
relation. In  particular, we show that  the B01 model is  only able to
reproduce  the  simulation  results  in  the galaxy  mass  range,  but
underpredicts the concentrations of  massive haloes.  On the contrary,
the  NFW  model  yields a  reasonable  fit  at  the massive  end,  but
dramatically underestimates the concentrations for haloes with $M \lta
10^{12}  \hMsun$.   We  present  a   new  model,  based  on  a  simple
modification  of the B01  model, that  that is  able to  reproduce the
concentration-mass relation  over the entire mass range  probed by our
numerical simulations.
\begin{table}
 \centering
 \begin{minipage}{140mm}
  \caption{Cosmological Parameters}
  \begin{tabular}{lcccccc}
\hline  Name &  $\Omega_{\Lambda}+\Omega_m$ & $\Omega_m$ & $h$ & $\sigma_8$ & $n$ & $\Omega_b$ \\
WMAP1  & 1.0 & 0.268 & 0.71 & 0.90 & 1.00 & 0.044 \\
WMAP3  & 1.0 & 0.238 & 0.73 & 0.75 & 0.95 & 0.042 \\
WMAP5  & 1.0 & 0.258 & 0.72 & 0.796 & 0.963 & 0.0438 \\
\hline 
\end{tabular}
\end{minipage}
\label{tab:params}
\end{table}

This  paper is organized  as follows:  in Section  \ref{sec:nbody} the
simulations  and   the  determination  of  the   halo  parameters  are
presented.     In    Sections    \ref{sec:cm},   \ref{sec:spin}    and
\ref{sec:shape} we discuss the  results for halo concentrations, spins
and  shapes respectively.  Finally,  Section~\ref{sec:conc} summarizes
the results.

\section{N-body simulations} 
\label{sec:nbody}

\begin{table}
 \centering
 \begin{minipage}{140mm}
  \caption{N-body Simulation Parameters}
  \begin{tabular}{lccccr}
\hline  Name &  Box  size  & N  &  part. mass  &  force  soft. & Nhalo\\
& $[{\rm Mpc}]$ &  & $[h^{-1}M_{\odot}]$  & $[h^{-1}{\rm kpc}]$ & $>500$ \\
\hline\\ 
W1-20.1   & 20  & $250^3$ & 1.36e7 & 0.43 & 814\\ 
W1-20.2   & 20  & $250^3$ & 1.36e7 & 0.43 & 849\\ 
W1-40.1   & 40  & $250^3$ & 1.09e8 & 0.85 & 1132\\ 
W1-40.2   & 40  & $250^3$ & 1.09e8 & 0.85 & 1018\\ 
W1-90.1   & 90  & $300^3$ & 7.19e8 & 1.92 & 2552\\ 
W1-90.2   & 90  & $300^3$ & 7.19e8 & 1.92 & 1992\\ 
W1-90.3   & 90  & $600^3$ & 8.98e7 & 0.85 & 12766\\ 
W1-180    & 180  & $300^3$ & 5.75e9 & 3.83 & 2406\\ 
W1-300    & 300  & $400^3$ & 1.12e10 & 4.72 & 6415\\
\hline\\                                   
W3-20.1   & 20   & $250^3$ & 1.32e7  & 0.43 & 960\\ 
W3-20.2   & 20   & $250^3$ & 1.32e7  & 0.43 & 1011\\ 
W3-40.1   & 40   & $250^3$ & 1.05e8  & 0.85 & 1108\\ 
W3-90.1   & 90   & $300^3$ & 6.94e8  & 1.92 & 1963\\ 
W3-90.2   & 90   & $300^3$ & 6.94e8  & 1.92 & 2043\\ 
W3-90.3   & 90   & $600^3$ & 8.67e7  & 0.85 & 13143\\ 
W3-180    & 180  & $300^3$ & 5.55e9  & 3.83 & 2105\\ 
W3-300    & 300  & $400^3$ & 1.08e10 & 4.78 & 5255\\ 
W3-360    & 360  & $400^3$ & 1.87e10 & 5.74 & 4947\\ 
W3-360.2  & 360  & $600^3$ & 5.55e9  & 3.83 & 17582\\ 
\hline\\
W5-20.1   & 20   & $250^3$ & 1.37e7  & 0.43 & 974\\ 
W5-40.1   & 40   & $250^3$ & 1.09e8  & 0.85 & 1119\\ 
W5-90.1   & 90   & $300^3$ & 7.21e8  & 1.92 & 1998\\ 
W5-180  & 180  & $300^3$ & 5.77e9  & 3.83 & 2302\\ 
W5-300  & 300  & $400^3$ & 1.13e10 & 4.74 & 5845\\ 
\hline 
\end{tabular}
\end{minipage}
\label{tab:sims}
\end{table}

Table  \ref{tab:sims}   lists  all   the  simulations  used   in  this
paper. Each simulation has a unique name, W$x$-LL.$n$, where $x=1,3,5$
refers to  the WMAP1, WMAP3  and WMAP5 cosmology, respectively,  LL is
the  simulation  box  size in  $\rm  Mpc$,  and  $n$ is  a  sequential
integer. For each cosmology we have run simulations for five different
box sizes,  which allows us to  probe halo masses  covering the entire
range $10^{10} \hMsun  \lta M \lta  10^{15} \hMsun$.  In
addition, in some case we  have run multiple (up to three) simulations
for the same  cosmology and box size, in order to  test for the impact
of cosmic  variance (and to increase  the final number  of dark matter
haloes).

All simulations have been performed  with PKDGRAV, a tree code written
by Joachim Stadel and Thomas Quinn (Stadel 2001). The code uses spline
kernel softening, for which  the forces become completely Newtonian at
2  softening lengths.   Individual time  steps for  each  particle are
chosen  proportional  to the  square  root  of  the softening  length,
$\epsilon$,    over   the   acceleration,    $a$:   $\Delta    t_i   =
\eta\sqrt{\epsilon/a_i}$. Throughout, we set $\eta = 0.2$, and we keep
the  value of the  softening length  constant in  co-moving coordinates
during each run. The physical values of $\epsilon$ at $z=0$ are listed
in  Table  \ref{tab:sims}.  Forces  are  computed using  terms  up  to
hexadecapole order  and a node-opening angle $\theta$  which we change
from $0.55$ initially  to $0.7$ at $z=2$.  This  allows a higher force
accuracy when the mass distribution  is nearly smooth and the relative
force errors can  be large. The initial conditions  are generated with
the GRAFIC2 package (Bertschinger 2001).  The starting redshifts $z_i$
are  set to  the  time when  the  standard deviation  of the  smallest
density fluctuations resolved within  the simulation box reaches $0.2$
(the smallest scale resolved  within the initial conditions is defined
as twice the intra-particle distance).
 
In  all  simulations,  dark  matter  haloes  are  identified  using  a
spherical overdensity (SO) algorithm.  Candidate groups with a minimum
of $N_f=250$ particles are selected using a FoF algorithm with linking
length $\phi  = 0.2 \times  d$ (the average particle  separation).  We
then:  (i) find  the point  $C$ where  the gravitational  potential is
minimum; (ii)  determine the radius $\bar  r$ of a  sphere centered on
$C$, where the density contrast  is $\Delta$, with respect to the {\it
  critical density}  of the Universe, $\rhocrit  =3H^2/8\pi G$.  Using
all  particles  in  the  corresponding  sphere we  iterate  the  above
procedure  until we  converge onto  a stable  particle set.   For each
stable particle set  we obtain the virial radius,  $\Rvir$, the number
of particles within  the virial radius, $\Nvir$, and  the virial mass,
$\Mvir$.   For our  adopted cosmologies  $\Delta\simeq  96.7$ (WMAP1),
$\Delta\simeq  93.5$ (WMAP3) and  $\Delta\simeq 95.1$  (WMAP5).  These
values are based on the fitting function of Mainini \etal (2003).

\begin{figure}
\psfig{figure=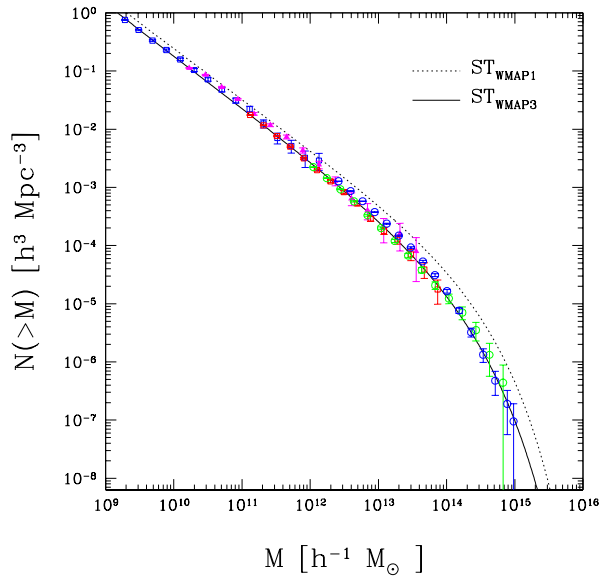,width=0.47\textwidth}
\caption{\scriptsize  Mass functions for  the WMAP3  simulations (only
  one simulation  per box  size is shown).  The doted and  solid lines
  represents the Sheth  \& Tormen (2001) prediction for  the WMAP1 and
  WMAP3 models respectively. Error bars indicate the Poisson noise in each mass bin.}
\label{fig:mf}
\end{figure}

Throughout this paper we only use haloes with $\Nvir>500$, of which we
there are  29944, 50177,  and 12238 in  the combined WMAP1,  WMAP3 and
WMAP5  simulations, respectively. For  completeness, Fig.~\ref{fig:mf}
shows  the halo  mass function  obtained from  the  WMAP3 simulations,
together  with the  predictions from  Sheth \&  Tormen (2001)  for the
WMAP1 and WMAP3 cosmologies (error bars show the Poisson noise in each mass bin). 
Note that the mass function obtained from
our simulation is in excellent agreement with this prediction and that
the WMAP3 cosmology predicts fewer haloes per co-moving volume than the
WMAP1 cosmology, especially at the massive end.

\subsection{Halo parameters}

For  each SO  halo in  our sample  we determine  a set  of parameters,
including the virial mass and radius, the concentration parameter, the
angular momentum, the spin  parameter and various axis ratios (shape).
Below  we  briefly  describe  how  these parameters  are  defined  and
determined. A more detailed discussion can be found in M07.

\subsubsection{Concentration parameter}

To compute the concentration of  a halo we first determine its density
profile. The halo center is defined  as the location of the most bound
halo particle, and  we compute the density ($\rho_i$)  in 50 spherical
shells, spaced  equally in logarithmic  radius. Errors on  the density
are  computed from  the  Poisson noise  due  to the  finite number  of
particles in  each mass shell.   The resulting density profile  is fit
with a NFW profile (Navarro \etal 1997):
\begin{equation}
\frac{\rho(r)}{\rhocrit} = \frac{\delta_{\rm c}}{(r/\rs)(1+r/\rs)^2},
\label{eq:nfw}
\end{equation}
During the  fitting procedure  we treat both  $\rs$ and  $\delta_c$ as
free  parameters.   Their values,  and  associated uncertainties,  are
obtained via a $\chi^2$  minimization procedure using the Levenberg \&
Marquart method. We define the r.m.s.  of the fit as:
\begin{equation}
\rhorms = \frac{1}{N}\sum_i^N { (\ln \rho_i - \ln \rho_{\rm m})^2}
\label{eq:rms}
\end{equation}
where $\rho_{\rm m}$ is the fitted NFW density distribution.
Finally,   we   define  the   concentration   of   the  halo,   $\cvir
\equiv\Rvir/\rs$,  using  the  virial  radius  obtained  from  the  SO
algorithm,   and    we   define   the    error   on   $\log    c$   as
$(\sigma_{\rs}/\rs)/\ln(10)$,  where  $\sigma_{\rs}$  is  the  fitting
uncertainty on $\rs$.   In order to compare our  results with previous
studies  it is  also  useful to  define, $c_{200}\equiv  r_{200}/\rs$,
where $r_{200}$ is the radius inside  of which the density of the dark
matter  halo  is 200  times  $\rhocrit$;  accordingly  we also  define
$M_{200}$ and $N_{200}$ as the mass and the number of particles within
$r_{200}$.

\subsubsection{Spin parameter}
\label{sec:spinpar}

The  spin  parameter is  a  dimensionless  measure  of the  amount  of
rotation of a  dark matter halo.  We use  the definition introduced by
Bullock \etal (2001b):
\begin{equation}
\lam=\frac{\Jvir}{\sqrt{2}\Mvir\Vvir\Rvir}
\end{equation}
where  $\Mvir,$  $\Jvir$  and  $\Vvir$  are the  mass,  total  angular
momentum  and circular velocity  at the  virial radius,  respectively. 
See  M07  for  a detailed  discussion  and  for  a comparison  of  the
different definitions for the spin parameter.

\subsubsection{Shape parameter}

Determining the shape of a three-dimensional distribution of particles
is a  non-trivial task (e.g.,  Jing \& Suto 2002).   Following Allgood
\etal (2006) we  determine the shapes of our  haloes starting from the
inertia tensor.  As  a first step, we compute the  halo's $3 \times 3$
inertia tensor using all the particles within the virial radius.  Next
we diagonalize the inertia tensor and rotate the particle distribution
according  to the  eigen vectors.   In this  new frame  (in  which the
moment  of inertia  tensor  is diagonal)  the  ratios $s=a_3/a_1$  and
$p=a_3/a_2$ (where $a_1 \geq a_2 \geq a_3$) are given by:
\begin{equation}
s \equiv {a_3 \over a_1} = \sqrt{ { \sum m_i z_i^2} \over \sum { m_i x_i^2}},
 \ \ \ \ \ \ \ \ p \equiv {a_3 \over a_2} = \sqrt{ { \sum m_i z_i^2} \over \sum { m_i y_i^2}}.
\end{equation}

Next we again compute the inertia tensor, but this time only using the
particles  inside the  ellipsoid defined  by $a_1$,  $a_2$ and  $a_3$. 
When deforming the ellipsoidal volume of the halo, we keep the longest
axis  ($a_1$) equal  to the  original radius  of the  spherical volume
($\Rvir$).  We  iterate this procedure  until we converge to  a stable
set of axis ratios.
  
\subsubsection{Offset parameter}

As in  M07, for  each halo we  compute the offset  parameter, $\xoff$,
defined as the  distance between the most bound  particle (used as the
center for the density profile) and the center of mass of the halo, in
units of the virial radius.  This offset is a measure of the dynamical
state of the  halo: relaxed haloes in equilibrium  will have a smooth,
radially symmetric density distribution,  and thus an offset parameter
that is virtually equal to zero.  Unrelaxed haloes, such as those that
have only  recently experienced a major  merger, are likely  to have a
strongly  asymmetric mass  distribution, and  thus a  relatively large
$\xoff$. Although some unrelaxed haloes  may have a small $\xoff$, the
advantage  of this  parameter  over, for  example,  the actual  virial
ratio, $2T/V$, as a function of radius (Macci\`o, Murante \& Bonometto
2003; Shaw \etal 2006), is that the former is trivial to evaluate.

\begin{figure*}
\psfig{figure=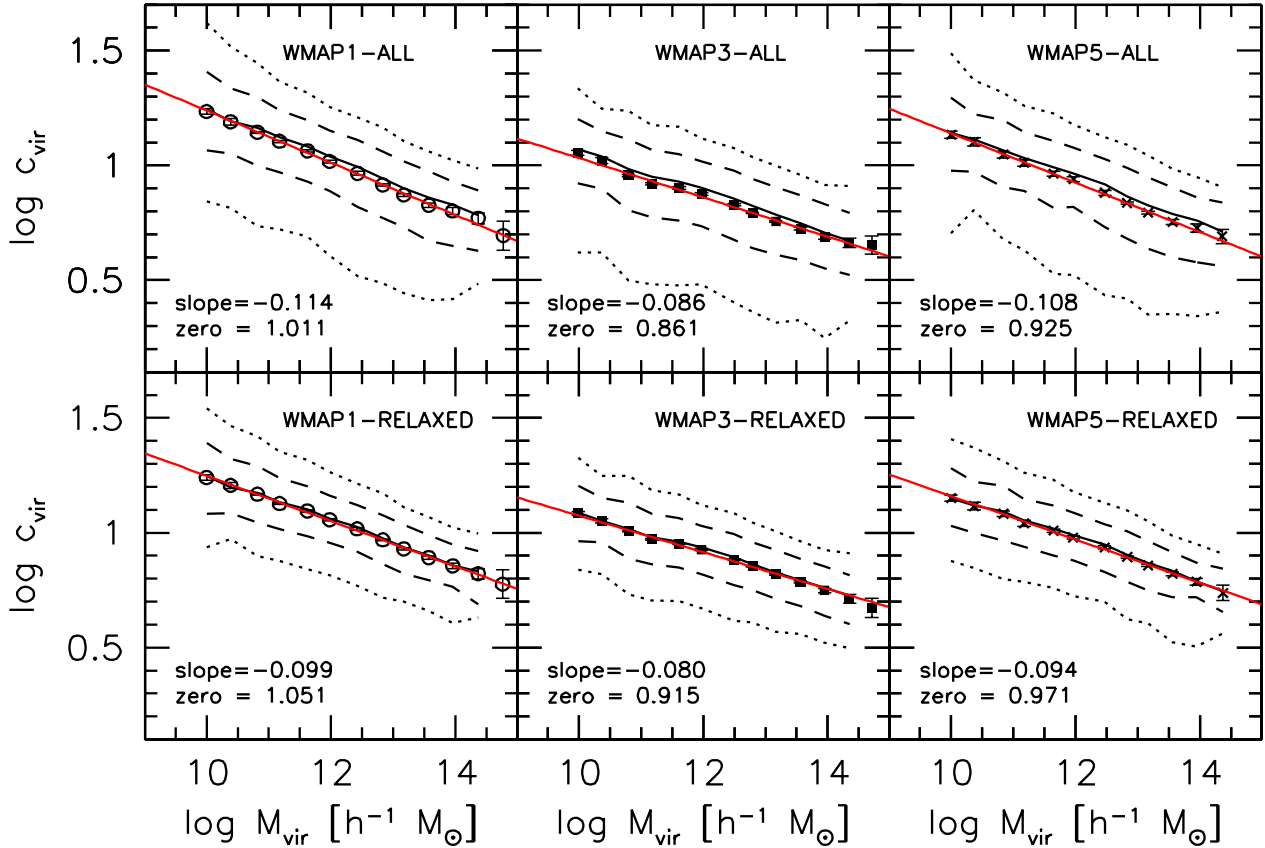,width=1.0\textwidth}
\caption{\scriptsize  $\cvir$  vs  $\Mvir$  for  WMAP1  (left),  WMAP3
  (center) and  WMAP5 (right).  The  upper panels show all  haloes with
  more than 500 particles within  $\Rvir$, while the lower panels show
  the ``relaxed'' haloes.  The  points show the mean concentration (in
  log $\cvir$) in  bins of width 0.4 dex in mass,  the error bar shows
  the Poisson  error on  the mean.  The  solid lines represent  the median
  concentration in each mass bin,  the dashed and dotted lines show the
  15.9,  84.1, 2.3 and  97.7th percentiles  of the  distribution.  The
  solid (red) line shows a power-law fit to the $\cvir-\Mvir$ relation:
  $\log  \cvir =  {\rm zero  + slope}(\log  \Mvir/12 h^{-1}M_{\odot})$
  whose parameters are  given in the lower left  corner of each panel,
  and in Table \ref{tab:fits1}.}
\label{fig:cm4s_vir}
\end{figure*}

\subsection{Relaxed Haloes}
\label{sub:relax}

Our halo  finder (and  halo finders in  general) does  not distinguish
between relaxed and  unrelaxed haloes.  There are many  reasons why we
might want  to remove unrelaxed halos.  First  and foremost, unrelaxed
haloes   often  have   poorly   defined  centers,   which  makes   the
determination  of   a  radial  density  profile,  and   hence  of  the
concentration parameter,  an ill-defined problem.   Moreover unrelaxed
haloes  often have  shapes that  are  not adequately  described by  an
ellipsoid, making our shape parameters ill-defined as well.

One could imagine  using $\rhorms$ (the r.m.s.  of the  NFW fit to the
density profile) to decide whether a halo is relaxed or not.  However,
while  it is  true  that  $\rhorms$ is  typically  high for  unrelaxed
haloes,  haloes  with  relatively  few  particles  also  have  a  high
$\rhorms$  (due to  Poisson noise)  even when  they are  relaxed  (cf. 
Fig.2  of M07  for the  correlation  between $\rhorms$  and $\Nvir$).  
Furthermore, since the spherical averaging used to compute the density
profiles has a  smoothing effect, not all unrelaxed  haloes have a high
$\rhorms$.  However,  these haloes are often characterized  by a large
offset  parameter,  $\xoff$.   We  therefore use  both  $\rhorms$  and
$\xoff$ to judge  whether a halo is relaxed or  not.  Following M07 we
split our halo sample in unrelaxed and relaxed haloes.  The latter are
defined as the haloes with $\rhorms  < 0.5$ and $\xoff < 0.07$.  About
70\% of  the haloes in our  sample qualify as relaxed  haloes. In what
follow, we will  present results for two different  samples of haloes:
ALL, which includes all haloes  with $\Nvir > 500$, and RELAXED, which
is the corresponding subsample of relaxed haloes.


\section{Concentration-Mass Relation}
\label{sec:cm}

\begin{figure*}
\begin{center}
\psfig{figure=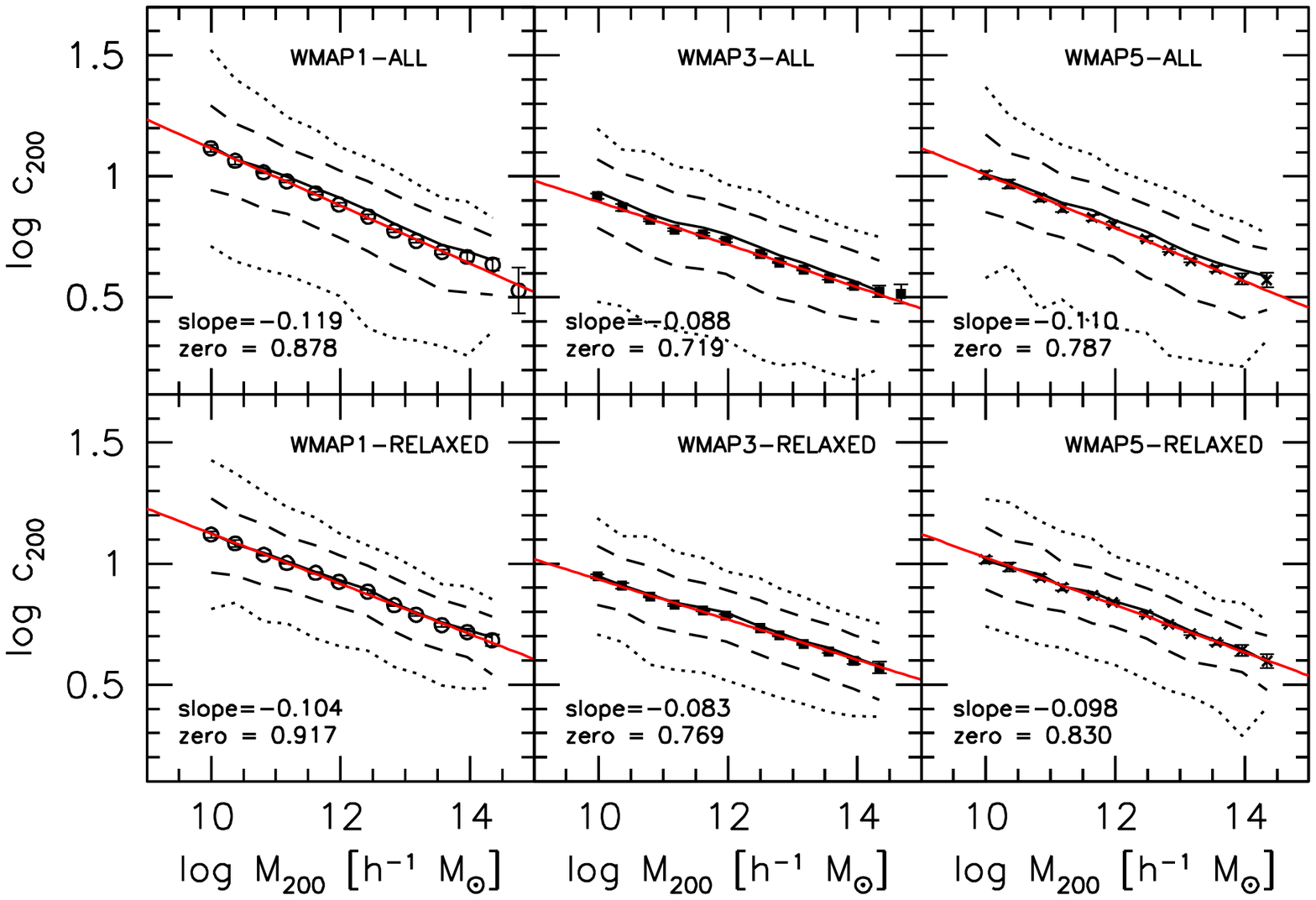,width=1.0\textwidth}
\caption{\scriptsize Same as Figure \ref{fig:cm4s_vir}, but for $c_{200}$ 
as function of $M_{200}$.}
\label{fig:cm4s_200}
\end{center}
\end{figure*}

\begin{figure}
\begin{center}
\psfig{figure=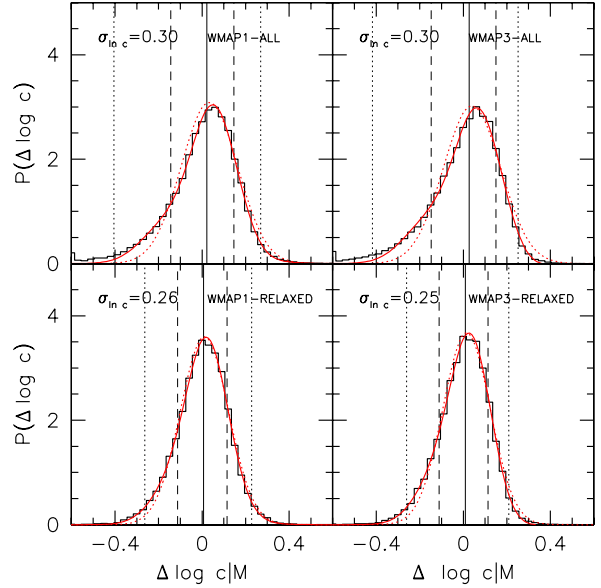,width=0.47\textwidth}
\caption{\scriptsize Histograms  of residuals from  the mean $c_{200}$
  mass relations  in Fig.~\ref{fig:cm4s_200}.  Only  results for WMAP1
  and  WMAP3  are  shown;  results  for  WMAP5  are  listed  in  Table
  \ref{tab:fits2}.  The  vertical lines show the  2.3th, 13.9th, 50th,
  84.1th, and 97.7th percentiles  of the concentration residuals.  The
  solid  red line  shows a  Gauss-Hermite polynomial  expansion  up to
  fourth order (whose parameters  are given in Table \ref{tab:fits2}),
  the dashed red  line shows the Gaussian corresponding  to the zeroth
  order of this expansion. The logarithmic variance ($\sigma_{\rm ln \
    c}$) of  the Gaussian fit  is reported in  the top left  corner of
  each panel.}
\label{fig:cm4hist}
\end{center}
\end{figure}

\subsection{Differences between WMAP1, WMAP3 and WMAP5}

We first discuss the  concentration-mass (hereafter $c$-$M$) relation. 
Figs.~\ref{fig:cm4s_vir} and~\ref{fig:cm4s_200} show the $\cvir-\Mvir$
and  $c_{200}-M_{200}$  relations  for  the  WMAP1,  WMAP3  and  WMAP5
cosmologies, and for  both the ALL and RELAXED  samples, as indicated. 
The  symbols show the  mean concentrations  (in logarithmic  space) in
mass bins  of 0.4~dex  width.  For all  three cosmologies  the $c$-$M$
relation is well fit by a  single power-law (red solid line).  For the
relaxed haloes, these best-fit power-laws relations are given by
\begin{equation}
\log \cvir = 1.051 -0.099 \log(\Mvir/[10^{12}h^{-1}M_{\odot}]) 
\end{equation}
\begin{equation}
\log c_{200} = 0.917 -0.104 \log(M_{200}/[10^{12}h^{-1}M_{\odot}])
\end{equation}
for  the  WMAP1   cosmology, and
\begin{equation}
\log \cvir = 0.915 -0.080 \log(\Mvir/[10^{12}h^{-1}M_{\odot}]) 
\end{equation}
\begin{equation}
\log c_{200} = 0.769 -0.083 \log(M_{200}/[10^{12}h^{-1}M_{\odot}]) 
\end{equation}
for the WMAP3 cosmology, and
\begin{equation}
\log \cvir = 0.971 -0.094 \log(\Mvir/[10^{12}h^{-1}M_{\odot}]) 
\end{equation}
\begin{equation}
\log c_{200} = 0.830 -0.098 \log(M_{200}/[10^{12}h^{-1}M_{\odot}]) 
\end{equation}
for the  WMAP5 cosmology. The  errors of these fitting  parameters are
listed in Table \ref{tab:fits1}.
Note that these relations differ in both the slope and the zero-point.
In particular, going from WMAP1  to WMAP5 to WMAP3, the slopes becomes
shallower and the zero-points become smaller.  Comparing the WMAP1 and
WMAP3 cosmologies, which are the extremes in terms of the cosmological
parameter values  (see Table \ref{tab:params}), the  difference in the
mean  $\log\cvir$ is 0.19,  0.15, 0.11  dex (i.e.,  a factor  of 1.55,
1.41,  1.29  in $\cvir$)  at  a  halo  mass of  $10^{10}$,  $10^{12}$,
$10^{14} h^{-1}M_{\odot}$.  A similar trend of lower normalization and
shallower  slopes is  also seen  going towards  higher redshift  for a
given cosmology  (e.g.  Zhao \etal  2003a).  This supports  the notion
that  the $c$-$M$  relation reflects  the assembly  histories  of dark
matter haloes: the  fact that WMAP3 haloes are  less concentrated than
their counterparts  in a WMAP1  (or WMAP5) cosmology,  simply reflects
that haloes assemble later in  a universe with lower $\Omega_m$ and/or
lower $\sigma_8$.  Note that the  three cosmologies also differ in the
spectral  index of  the  matter  power spectrum,  $n$,  which is  also
responsible for some of the differences in the $c$-$M$ relations.

Fig.~\ref{fig:cm4hist}  shows the  distributions of  the concentration
residuals, $\Delta \log c_{200}$,  relative to the best-fit power-laws
$c_{200}(M_{200})$, for the WMAP1 and WMAP3 cosmologies.  The full set
of  haloes   (upper  panels)  shows  a  clear   skewness  towards  low
concentrations.   This was  already pointed  out by  M07, and  is even
apparent in the simulations of B01.  As discussed by M07, this tail of
low  concentration parameters  is due  to unrelaxed  haloes.  Removing
these haloes results in  almost Gaussian distributions in $\Delta \log
c_{200}$,    as   is    apparent    from   the    lower   panels    in
Fig.~\ref{fig:cm4hist}.   Table  \ref{tab:fits2}  lists  a  number  of
parameters  for these  distributions,  including those  for the  WMAP5
cosmology  (not  shown  in  Fig.~\ref{fig:cm4hist}).   Note  that  the
scatter  these  distributions is  virtually  the  same  for all  three
cosmologies; hence,  although the slope and zero-point  of the $c$-$M$
relation is clearly cosmology dependent, the scatter is not.

\subsection{Comparison of WMAP1 results with the Millennium Simulation}

\begin{figure*}
\begin{center}
\psfig{figure=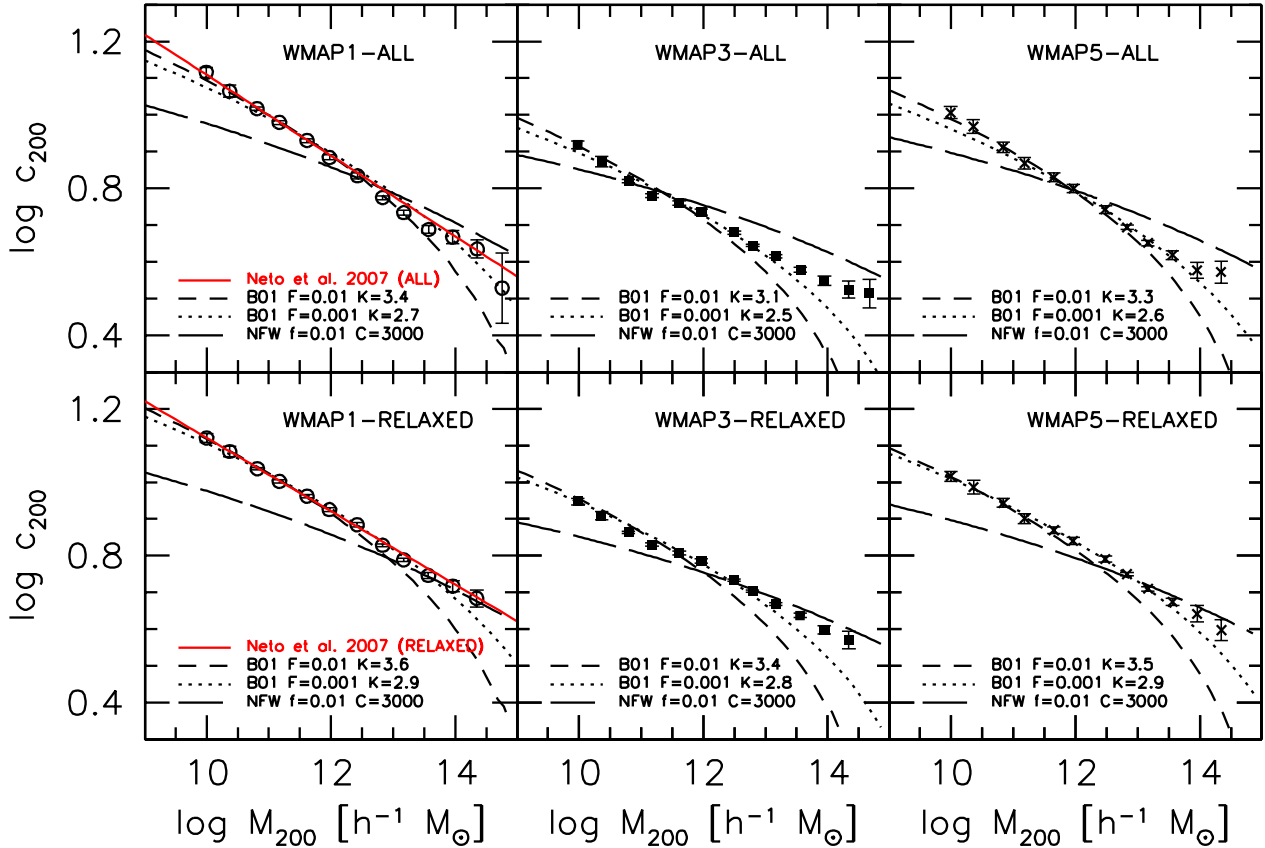,width=1.0\textwidth}
\caption{\scriptsize   Comparison    between   the   $c_{200}-M_{200}$
  relations in our  simulations and those predicted by  the toy models
  of Bullock  \etal (2001a,  dotted and dashed  lines) and  NFW (1997)
  (long-dashed lines). For the Bullock  \etal models we show the value
  of $K$ that results in the  best fit for galaxy mass haloes, for the
  NFW model  we use the  same value of  $C$ for all models.  The solid
  (red)   line  shows  the   concentration-mass  relations   from  the
  Millennium simulations  as measured by Neto \etal  (2007).  See text
  for further details.}
\label{fig:cm4_200_2}
\end{center}
\end{figure*}

Recently,  Neto \etal  (2007) analysed  the $c$-$M$  relation  of dark
matter  haloes in  the  Millennium simulation  (Springel \etal  2005).
They confirm many of the results published previously in M07, namely:
\begin{enumerate}
\item The concentration mass relation (at redshift zero) is well
  described by a single power-law with slope $\simeq -0.10$.
\item The inclusion of un-relaxed haloes biases the zero-point low,
  and increases the scatter.
\item The  inclusion of un-relaxed  haloes is largely  responsible for
  creating a false correlation  between concentration parameter  and spin parameter.
\end{enumerate}
The  symbols  in Fig.~\ref{fig:cm4_200_2}  show  the  same mean  $\log
c_{200}$ as function of $M_{200}$ as shown in Fig.~\ref{fig:cm4s_200}.
The  solid red  lines in  the left-hand  panels indicate  the best-fit
relation   obtained  by   Neto  \etal   (2007)  from   the  Millennium
simulation\footnote{Note that the cosmology adopted for the Millennium
  simulation is very close, but not  exactly the same as for our WMAP1
  cosmology.   For  the  purpose  of  our  discussion,  though,  these
  differences are completely negligible.}. For both the full sample of
haloes and for the subsamples of relaxed haloes, the Neto \etal (2007)
results are in  excellent agreement with our simulation  data over the
range of haloes probed our simulations. Note that, for comparison, the
Millennium  simulation only  covered  the range  $10^{12} \hMsun  \lta
M_{200} \lta 10^{15}$ $\hMsun$.  This agreement is extremely encouraging
because the  Millennium simulation was  run with a  different $N$-body
code  (GADGET, rather  then PKDGRAV),  and  Neto \etal  (2007) used  a
different halo finder.

Yet, Neto \etal (2007) questioned  the results of M07 at small masses.
They  argued that  at  least  1000 particles  are  needed to  reliably
determine halo concentrations, and  they conjecture that the small box
size  (20 Mpc) of  the highest  resolution simulation  in M07  will be
biased by missing  large scale power.  They conclude  that some of the
differences found in M07 between  the Eke \etal (2001) model and their
simulation results  at small  scales (at a  few $10^{10}  \hMsun$) are
unreliable.  Note, however, that the differences between the Eke \etal
(2001) model  and the  M07 simulations are  statistically significant,
even at  $10^{11}\hMsun$ (i.e on  scales resolved with more  than 1000
particles). Furthermore, the speculation by Neto \etal (2007) that the
20~Mpc boxes give biased halo  concentrations is not justified: as was
clearly  shown  in  M07,  the  scatter  in  the  $c$-$M$  relation  is
independent of  the large scale  environment.  To further  bolster the
results of M07 at masses below $1\times 10^{11} \hMsun$, in this paper
we include new simulations of 90 Mpc boxes with 216 million particles.
These  simulations have  a particle  mass of  $9 \times  10^7 \hMsun$,
which  is an order  of magnitude  higher than  that of  the Millennium
simulation.

\begin{figure*}
\begin{center}
\psfig{figure=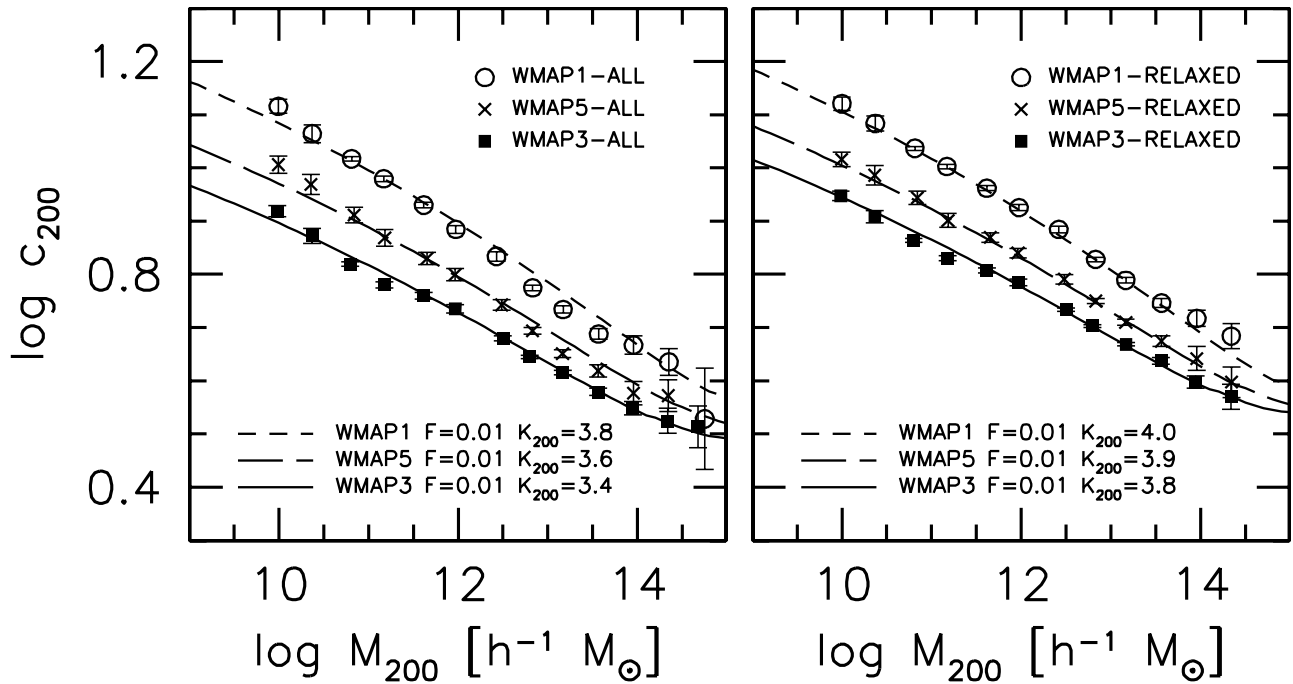,width=1.0\textwidth}
\caption{\scriptsize   Comparison    between   the   $c_{200}-M_{200}$
  relations  in our  simulations and  those predicted  by  our revised
  version of the  B01 model.  Our new model  is able to simultaneously
  fit both the  low mass and high mass end of  the $c-M$ relation.  It
  provides an excellent fit to  the simulations results over more than
  4 orders of magnitude in halo mass.}
\label{fig:cm4_200_3}
\end{center}
\end{figure*}

\subsection{Comparison with toy models}

A number of  studies have presented analytical models  for the $c$-$M$
relation, that have been calibrated against numerical simulations, but
only over a relatively narrow range  in halo mass. Here we compare our
simulation results, which cover 5 orders of magnitude in halo mass, to
two of  the most commonly used  models: that by  Bullock \etal (2001a;
hereafter the B01  model), and that by Navarro  \etal (1997; hereafter
the NFW model). Both these models  have 2 free parameters ($F$ and $K$
for B01, $f$  and $C$ for NFW) that have been  tuned to reproduce some
simulation  results.  Fig.~\ref{fig:cm4_200_2}  shows  the predictions
for the $c-M$  relation for these models.  B01  advocated a model with
$F=0.01$ and $K=4.0$  which fit the $c-M$ relation  of galaxy sized DM
haloes ($10^{11}-10^{13} \hMsun$).  These parameters were obtained for
a flat $\Lambda$CDM cosmology  with $\Omega_m=0.3$ and $\sigma_8=1.0$. 
A slightly  lower normalization, $K=3.4$,  was advocated by M07  for a
WMAP1 cosmology (see  Zhao \etal 2003a and Kuhlen  \etal 2005 for more
details).

This $K=3.4,  F=0.01$ model is  shown as the  long dashed line  in the
upper left-hand panel (ALL  sample) of Figure \ref{fig:cm4_200_2}.  It
fits  the data  from  $M=10^{10} \hMsun$  to  $M=10^{13} \hMsun$,  but
under-predicts the  $c-M$ relation at  high masses. Neto  \etal (2007)
argue that this is a fundamental failure of the B01 model. However, as
stressed by B01  (in both their paper and  in their publicly available
source code), their  model is not expected to work  on scales where $F
\times  \Mvir \gta M_{*}$.   A partial  solution is  to adopt  a lower
value for $F$; for instance  using $F=0.001$ and $K=2.7$ yields a good
fit to our simulation data over  the range $10^{11} \hMsun \lta M \lta
10^{14} \hMsun$ (dotted line).

By contrast, the NFW model is in reasonable agreement with the $c$-$M$
relation of  relaxed haloes  at the high  mass end.  This  model, with
parameters  advocated in the  original paper  by Navarro  \etal (1997)
($f=0.01,  C=3000$), is given  by the  long-dashed lines.   This model
matches  the  slope of  the  $c-M$  relation  for the  highest  masses
$M>10^{13}  \hMsun$.   However,   it  dramatically  underpredicts  the
concentrations of  low mass haloes; in  fact, the NFW  model predict a
slope for  the $c$-$M$ relation which  is much shallower  than what we
infer from  our simulations, causing the discrepancy  to increase with
decreasing halo mass.  Neto  \etal (2007) gloss over these differences
by saying  that the differences are  small compared to  the scatter in
$c$,  i.e.   since  the  scatter  in  $c$ is  a  factor  of  1.4,  the
discrepancy between the  NFW model and the simulation  results is less
than 1$\sigma$.   However, the relevant  parameter to compare  with is
the error  on the  mean (or  median) $c$.  Given  the numbers  in Neto
\etal (2007) we  obtain a Poisson error of 0.004  dex (i.e.  less than
1\%) for  their lowest  mass bin around  $10^{12} \hMsun$.   We obtain
similarly  small errors with  our data  for haloes  of the  same mass.
Thus,  we argue  that  despite the  large  amount of  scatter in  halo
concentrations  at fixed halo  mass, the  discrepancy between  the NFW
model and  the simulation results are very  significant: at $M=10^{12}
\hMsun$ we find the the NFW model underpredicts the mean concentration
by a factor 1.12 (a discrepancy  of 0.05 dex in $\log c_{200}$), which
increases  to  a  factor of  1.38  (0.14  dex  in $\log  c_{200}$)  at
$M=10^{10}  \hMsun$.   Of course  systematic  errors  in fitting  halo
concentrations may well be larger  than 1\%, which would cause them to
dominate the error  budget.  However, the fact that  Neto \etal (2007)
obtain results  that are virtually  identical to us using  a different
$N$-body  simulation code, a  different halo  finder, and  a different
fitting  procedure,  suggests that  these  effects  are  not going  to
mitigate the problem of the NFW model at low masses.

Both  the B01  and NFW  models predict  that the  $c$-$M$  relation is
shallower for  the WMAP3  cosmology than for  the WMAP1  cosmology, in
quantitative  agreement  with the  simulation  results.  However,  the
problems at  high masses for the B01  model and at low  masses for the
NFW  models remain.   Another problem  for  these models  is that  the
normalization  of  the  concentration   mass  relation  in  the  WMAP3
cosmology is lower  than expected based on the  parameters $K$ for B01
and $C$ for  NFW that are required to match the  $c-M$ relation in the
WMAP1 cosmology. Since the B01 model does not fit the $c$-$M$ relation
well at  high masses, a  straight forward $\chi^2$  minimization gives
values of $K$  biased high. Thus, to determine  the best-fit values of
$K$, we  construct a grid of  models with $K$ varying  at intervals of
0.1, and chose  the value of $K$ which provides the  best fit (by eye)
over the widest  range of halo masses. For $F=0.01$  the range in mass
that is  well fitted is  $M\sim 10^{10-12}\hMsun$, while  for $F=0.001$
the range in mass is $M\sim 10^{11-13}\hMsun$.

\subsection{A revision of the Bullock model}

We  now discuss  a  modification of  the  B01 model  that retains  the
agreement for galaxy mass haloes, but improves the agreement for group
and cluster sized haloes.

Both the  B01 and NFW  models are based  on the idea that  the central
densities  of dark  matter  haloes  reflect the  mean  density of  the
universe at a  time when the central region of  the halo was accreting
matter at  a high  rate.  Therefore haloes  with central  regions that
collapse earlier are  expected to be denser than  haloes that collapse
later. Thus the first  step in the B01 (and NFW) model  is to assign a
redshift of  collapse, given a halo  of mass $\Mvir$ at  a redshift of
observation, $z$.

B01 define the collapse redshift,  $\zc$, as the redshift at which the
characteristic mass is equal to a fraction $F$ of the halo mass at the
observation redshift, $z$,
\begin{equation}
M_*(\zc)=F \, \Mvir(z)\,,
\end{equation}
Using the  spherical collapse  formalism, this characteristic  mass is
defined via
\begin{equation}
\sigma(M_*,\zc)=\sigma(M_*,0) D(\zc)=1.686
\end{equation}
where $\sigma(M,z)$ is the rms overdensity on mass scale $M$ at redshift
$z$, and $D(z)$ is the linear growth rate.  

For comparison,  NFW define the  collapse redshift as the  redshift at
which,  according  to  the  Press  \& Schechter  formalism  (Press  \&
Schechter 1974;  Lacey \& Cole 1993),  half of the virial  mass of the
halo was first  contained in progenitors more massive  than a fraction
$f$ of the final mass.

The next step in the B01 (and NFW) model is to link the density of the
halo  at $z$  to  that of  the mean  density  of the  universe at  the
collapse redshift.  First we define the mass of the halo via
\begin{equation}
\Mvir(z) = { 4 \over 3} \pi \Rvir^3(z) \Deltavir(z) \rhou(z),
\end{equation}
where $\Deltavir$ is the overdensity  of the halo relative to the {\it
  mean density}  of the universe, $\rhou$.  Alternatively  one can set
$\Deltavir(z)\rhou(z)=\Delta(z)\rhocrit(z)$,  where  $\Delta$  is  the
overdensity of the halo with  respect to the {\it critical density} of
the  universe, $\rhocrit$.  A  common choice  is to  set $\Delta=200$,
which  results in  the  definition  of virial  mass  and radius  being
independent   of  cosmology   (using  units   of  $h=1$).    Then  the
characteristic density  of the halo at  any epoch (as  defined by B01)
can be written as:
\begin{equation}
  \tilde{\rhos}(z)= { {3 \Mvir(z)} \over {4 \pi \rs^3(z)}}
= \cvir^3(z) \Deltavir(z) \rhou(z),
\end{equation}
where  $\cvir(z)=\Rvir(z)/\rs(z)$.   
 
B01  identified  the  characteristic   density  of  the  halo  at  the
observation redshift,  $z$, with the  mean density of the  universe at
the collapse redshift via
\begin{equation}
  \tilde{\rhos}(z)=K^3 \Deltavir(z)\rhou(\zc)
\end{equation}
where  $K$ is  a  constant  to be  determined  by calibration  against
numerical  simulations.  Since $\rhou(\zc)=\rhou(z)(1+\zc)^3/(1+z)^3$,
the concentration at the observation redshift is given by
\begin{equation}
\cvir(\Mvir,z) = K (1+\zc)/(1+z).
\end{equation}

Our  modification  to the  B01  model is  to  simply  assume that  the
characteristic density  of the halo $\tilde{\rhos}$  is independent of
redshift (i.e. $\tilde{\rhos}(z)=\tilde{\rhos}(\zc)$).  Thus
\begin{equation}
\cvir(\Mvir,z) = K \, 
\left[\frac{\Deltavir(\zc)}{\Deltavir(z)}\frac{\rhou(\zc)}{\rhou(z)}\right]^{1/3}
  \equiv K g(\zc,z).
\end{equation}
As for the original model,  the (free) parameter $K$, which represents
the concentration of  the halo at the collapse  redshift $\zc$, has to
be calibrated against  numerical simulations.  The function $g(\zc,z)$
specifies the  growth of the halo concentration  between the redshifts
of collapse  and observation.  Note  that this equation is  similar to
that  of the  Eke  \etal  (2001) model,  except  that they  implicitly
assumed that  $\cvir(\zc)=1$ (which is  wrong), and they  define $\zc$
differently.

We can also express $g$ in terms of $\Delta(z)$ and $\rhocrit$
\begin{equation}
  g(\zc,z)= \left[\frac{\Delta(\zc)}{\Delta(z)}\frac{\rhocrit(\zc)}{\rhocrit(z)}\right]^{1/3}
\end{equation}
In the case of the $\Delta=200$ halo definition, we thus have that
\begin{equation}
c_{200}(z)=K_{200} \left[ \frac{\rhocrit(\zc)}{\rhocrit(z)}\right]^{1/3} 
= K_{200} \left[ \frac{H(\zc)}{H(z)} \right]^{2/3}.
\end{equation}
So that  the evolution in $c_{200}$  is given by the  evolution of the
Hubble parameter, which is given by:
\begin{equation}
{H^2(z) \over H^2_0} = \left[ \Omega_\Lambda + \Omega_m(1+z)^3 + 
(1-\Omega_\Lambda -\Omega_m)(1+z)^2\right],.
\end{equation}

Our modification of  the B01 model only consists  of taking account of
the redshift dependence of the  halo density contrast.  In the case of
the   $\Delta=200$  halo  definition,   the  redshift   dependence  of
$\Deltavir=\Delta/\Omega_m$   is  completely   governed  by   that  of
$\Omega_m$.  For an EdS  universe $\Omega_m(z)=1$, so that our revised
model yields  exactly the same  $\cvir(z)$ as the original  B01 model.
However, for a \LCDM universe $\Omega_m$ is a function of redshift, so
that the  difference between $\Omega_m(\zc)$ and  $\Omega_m(z)$ can be
significant, and  thus also the  difference in $\cvir(z)$  between the
B01 model and our revised version.

\begin{figure}
\psfig{figure=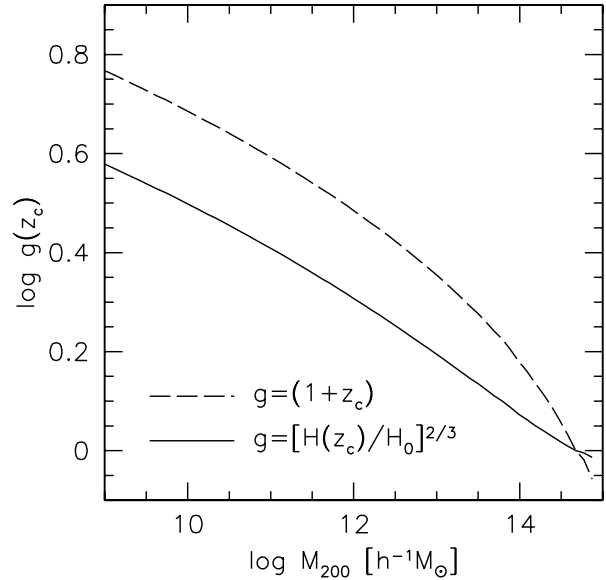,width=0.47\textwidth}
\caption{\scriptsize  Growth factor  of concentration  parameter  as a
  function of halo  mass (computed at $z=0$) for  WMAP1 cosmology. The
  dashed line  shows the  growth factor for  the B01 model,  while the
  solid line shows  the growth factor for our  revised B01 model.  Our
  model has a shallower mass  dependence at high masses ($\gta 10^{13}
  \hMsun$), resulting in a better  agreement with the high mass end of
  the concentration mass relation (cf. Fig. \ref{fig:cm4_200_3}).}
\label{fig:gz}
\end{figure}

\begin{figure}
\psfig{figure=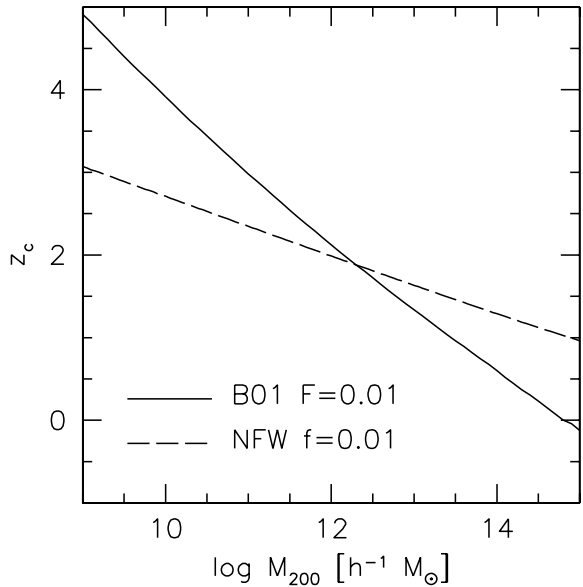,width=0.47\textwidth}
\caption{\scriptsize Formation  redshifts as  a function of  halo mass
  for the  B01 and NFW models.  The  NFW model has a  much weaker mass
  dependence  to the  formation  redshift than  the  B01 model,  which
  results in a $c-M$ relation much shallower than in the simulations
  as shown in Figure \ref{fig:cm4_200_2}.}
\label{fig:zf}
\end{figure} 

Fig.\ref{fig:gz} compares  the concentration growth  factors, $g(\zc)$
for the B01 model ($g(\zc) =  (1+\zc)$) and for our new model ($g(\zc)
=  [H(\zc)/H_0]^{2/3}$)  in  a  WMAP1  cosmology.   For  haloes  below
$M=10^{12}  \hMsun$  (and collapse  redshifts  $>  2$)  the slopes  of
$g(M_{200})$ are the same for both models.  This owes to the fact that
$\Omega_m(z) \rightarrow  1$ at  high redshifts.  However  for massive
haloes (and  low collapse redshifts), the  concentration growth factor
$g$  has a  shallower slope  in  our model  than in  the original  B01
formulation.   Because  of  this  difference our  new  model  predicts
concentrations at  the high mass  end that are relatively  higher than
for the original  B01 model.  Note also that at the  low mass end, the
new  concentration growth  factor is  lower  than for  the B01  model;
consequently, in order  to match the same $c$-$M$  relation at the low
mass end, the $K$ parameter in  our new formulation has to be somewhat
higher than in the B01 model.

The  symbols  in Fig.~\ref{fig:cm4_200_3}  show  the $c$-$M$  relation
obtained from  our simulations  for all three  cosmologies, separately
for  all haloes  (left-hand panel)  and for  the subsample  of relaxed
haloes  (right-hand  panel).  The  lines  correspond  to the  best-fit
$c$-$M$ relations  obtained using our new  toy model, in  which we let
$K$ be  a free parameter.  The  best-fit values of $K$  are indicated. 
Note that  our new model can  fit the simulation  data remarkably well
over  the  entire range  of  halo masses  probed,  and  for all  three
cosmologies. Clearly, this new model is a significant improvement over
the B01  and NFW models.  As with  the NFW model and  the original B01
model,   however,   different    cosmologies   require   a   different
normalization parameter $K$. In  particular, the best-fit value of $K$
decreases going from WMAP1 to WMAP5 to WMAP3.  This is unfortunate, as
there is currently  no existing model that can  {\it a priori} predict
the  value  of  $K$  given  the  cosmology.   Consequently,  for  each
cosmology, the best-fit value of  $K$ has to be determined empirically
using   high-resolution   numerical   simulations.   For   the   three
cosmologies considered here, the differences in the best-fit values of
$K$ are small ($\simeq 5\%$),  but they nevertheless indicate that our
model  does not  fully capture  the power  spectrum dependence  of the
$c-M$ relation (nor  does the original B01 model.   We plan to address
this issue in more detail in a future paper.


\subsection{Why does the NFW model fail?}

There are  two differences between the  NFW model on the  one hand and
the B01  model and  it modification suggested  here on the  other: the
definition of halo formation redshift, and the method used to link the
mean  density  of  the  universe  at the  formation  redshift  to  the
concentration   of   the   halo   at   the   observation   redshift.   
Fig,~\ref{fig:zf} shows the formation  redshifts as a function of halo
mass for the  NFW and B01 models in a WMAP1  cosmology.  The NFW model
predicts  formation  redshifts  with  a significantly  shallower  mass
dependence  that the  B01 model.   We have  experimented with  a model
using the NFW formation time,  but the B01 method of linking formation
redshift to  halo concentration.  For  none of these  models, however,
were we able  to obtain a concentration-mass relation  with a slope as
steep as found in the  simulations. Consequently, we conclude that the
main problem with the NFW model is its definition of the halo collapse
redshift.

\subsection{Comparison with observations}

Several studies have addressed the issue of the apparent inconsistency
between the inner  density slopes of dwarf and  LSB galaxies and those
predicted by  CDM (e.g., Moore 1994;  Flores \& Primack  1994; van den
Bosch \& Swaters  2001; de Blok \etal 2001,  2002; Swaters \etal 2003,
Dutton  \etal 2005;  Gentile \etal  2005; Simon  \etal 2005;  Kuzio de
Naray \etal 2008).
For  more massive spiral  galaxies the  situation is  more complicated
because the  baryons contribute significantly (and  may even dominate)
the  total  mass profile  in  the inner  regions,  and  thus pure  CDM
predictions are  difficult to be  tested.  For these reasons  we limit
our comparison to data of dwarf and LSB galaxies.
Measuring  inner density slopes  of dark  matter haloes  from rotation
curves is difficult.   A common practice is to fit  a power-law to the
rotation curve,  either to  the inner few  data points (e.g.,  de Blok
2001); or to  the full rotation  curve (e.g. Simon \etal  2005).  Both
methods have some drawbacks. First of all, the inner few points of the
rotation  curve carry  with them  the largest  systematic  errors.  In
addition, realistic dark matter halo density profiles (whether they be
pseudo-isothermal,  NFW,   or  generalized  NFW   profiles)  all  have
continuously varying  density slopes with radius, so  that the results
of fitting  a single  power-law are extremely  sensitive to  the exact
radial  range covered.   A more  robust approach  is to  fit  a double
power-law density profile to the  entire rotation curve (e.g., van den
Bosch \& Swaters 2001, Dutton  \etal 2005).  However, as emphasized by
these  studies,  even in  this  case  there  are certain  degeneracies
inherent  to  the mass  modeling  that  make  it difficult  to  obtain
stringent constraints on  the inner density slopes, even  with data of
high spatial resolution.

In addition to predictions for the slope of the inner density profile,
the CDM  model also makes  predictions regarding the  concentration of
dark  matter  haloes.   Unfortunately, measuring  halo  concentrations
requires knowledge of the virial radius of the halo, a parameter which
is poorly constrained from the  observations.  This is due to the fact
that a  rotation curve only probes  the inner $\sim 10\%$  of the dark
matter halo.

An observationally  more robust measurement of the  central density is
given by  the dimensionless  parameter $\Delta_{V/2}$, defined  as the
average density  of the  halo, with respect  to the  critical density,
inside the radius $R_{V/2}$, where the halo circular velocity drops to
half its maximum value (Alam, Bullock \& Weinberg 2002).
\begin{equation}
\Delta_{V/2} = \frac{\bar{\rho}(R_{V/2})}{\rhocrit} = 
50 \left[ \frac{\Vmax}{\kms} \right ]^2 \left[ \frac{h^{-1} \kpc}{R_{V/2}} \right ]^2
\end{equation}

For an NFW halo, the  circular velocity reaches a maximum, $\Vmax$, at
a radius  $r \simeq 2.163  \,\rs$, so that  $R_{V/2} \sim 0.126  \rs$. 
The maximum circular velocity is given by
\begin{equation}
\Vmax^2 = 0.2162 \, \Vvir^2 \,\cvir/f(\cvir),
\end{equation}
where $f(x) = \ln(1+x) -x/(1+x)$. Note that the factor 0.2162 is equal
to $f(x)/x$,  with $x = 2.163$.   The virial velocity  scales with the
virial mass and virial radius according to
\begin{equation}
\frac{\Vvir}{\kms} = G^{1/3}\left[\frac{\Mvir}{\hMsun}\right]^{1/3} \left[ \frac{\Delta}{200}\right]^{1/6} 
\end{equation} 
\begin{equation}
\frac{\Vvir}{\kms} = \frac{\Rvir}{h^{-1}\kpc} \left[\frac{\Delta}{200}\right]^{1/2}.
\end{equation} 
with $G$  the gravitational constant.   Using these relations  and the
definition   of  the   concentration   parameter  ($\cvir=\Rvir/\rs$),
$\Delta_{V/2}$ can  be expressed purely in terms  of the concentration
parameter:
\begin{equation}
\Delta_{V/2} = 680.9 \frac{\Delta}{200} \frac{\cvir^3}{f(\cvir)}.
\end{equation}
Thus given a  relation between $\cvir$ and $\Mvir$  for NFW haloes, we
can convert this into a relation between $\Delta_{V/2}$ and $\Vmax$.

\begin{center}
\begin{figure}
\psfig{figure=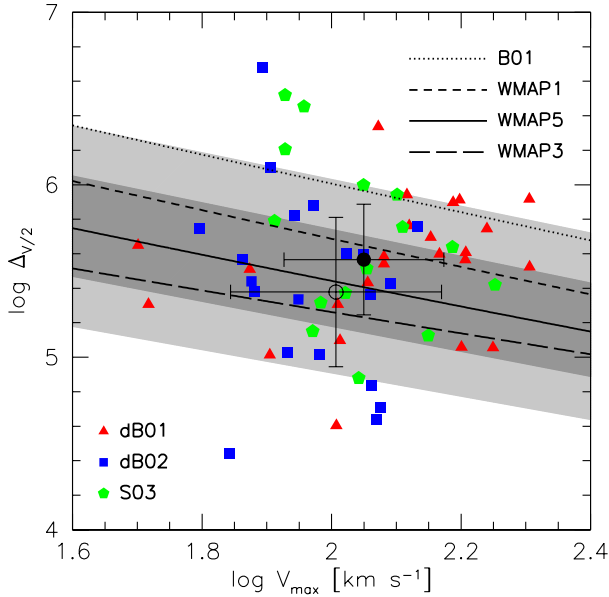,width=0.47\textwidth}
\caption{\scriptsize  Comparison  between  the  $\Delta_{  V/2}-\Vmax$
  relation from  observations of dwarf and LSB  galaxy rotation curves
  and  the predictions  of N-body  simulations in  a variety  of \LCDM
  cosmologies.  The observations are: de Blok, McGaugh \& Rubin (2001,
  dB01); de Blok \& Bosma (2002, dB02); and Swaters \etal (2003, S03).
  The   shaded  region   shows  the   68.3\%  and   95.4\%   range  of
  $\Delta_{V/2}$  from our  WMAP5  simulations. See  text for  further
  details.}
\label{fig:dv2}
\end{figure}
\end{center}

Fig.~\ref{fig:dv2}  shows  the  relation  between  $\Delta_{V/2}$  and
$\Vmax$  for  both observations  (symbols)  and  theory (lines).   The
observations are for dwarf and  LSB galaxies, color coded according to
the   reference.    These  values   are   calculated   based  on   the
pseudo-isothermal halo  fits to the observed rotation  curves. We have
removed  a few galaxies  with obviously  very bad  fits, or  for which
$\Vmax$ was  poorly constrained  by the data.   We have  verified that
including these galaxies makes no significant difference to the median
observed $\Delta_{V/2}$.   The solid black circle shows  the median of
the  data points,  and the  error bars  reflect the  median  errors on
$\Delta_{V/2}$ and  $\Vmax$. Although  these galaxies are  dark matter
dominated,  the  baryons make  a  non-negligible  contribution to  the
rotation curve.   The open circle shows the  median $\Delta_{V/2}$ and
$\Vmax$ for the dark halo  when the contribution of the baryons (stars
and gas)  to the rotation  curve has been subtracted.   These baryonic
contributions  are available  for about  half of  the  galaxies.  This
results in  a $\Delta_{V/2}$ that  is approximately 0.16 dex  (i.e., a
factor of  1.45) lower, and a  $\Vmax$ that is $\sim  0.04$ dex (i.e.,
factor of  1.10) lower.   We note that  the median  $\Delta_{V/2}$ and
$\Vmax$ without subtracting  the baryons is the same  for this subset,
as  the full  sample, so  that the  differences in  $\Delta_{V/2}$ and
$\Vmax$ are not simply caused by a selection effect.

The  lines  show  the   predictions  for  $\Delta_{V/2}-\Vmax$  for  4
cosmologies.  The dotted  line shows the \LCDM cosmology  used in Alam
\etal   (2002),  which  has   $\Omega_m=0.3$,  $\Omega_{\Lambda}=0.7$,
$\sigma_8=1.0$,  $h=0.7$ and $n=1$.   As noted  by Alam  \etal (2002),
this cosmology  results in  central densities that  are a factor  of 4
higher  than   observed.   However,  this  cosmology   has  values  of
$\sigma_8$,  $\Omega_m$ and  $n$  that are  significantly higher  than
those    for   the    WMAP   cosmologies    considered    here.    The
$\Delta_{V/2}-\Vmax$ relations for  WMAP1, WMAP3 and WMAP5 cosmologies
are  shown   as  the  short  dashed,  long-dashed   and  solid  lines,
respectively.   Note that  the data  are broadly  consistent  with all
these  cosmologies; within  the  errors,  the data  seem  to prefer  a
cosmology with $\sigma_8\simeq 0.8$ and $n\simeq 0.96$.

Note,  though, that there  are several  systematic effects  that could
alter  this conclusion.   If  the  effect of  halo  contraction (e.g.  
Blumenthal  \etal 1986) is  taken into  account, it  will result  in a
lower  $\Delta_{V/2}$  for  the  initial  halo  and,  and  thus  favor
cosmologies  with lower  $\sigma_8$.   On the  other  hand, there  are
several systematic  effects that  result in underestimates  of $\Vmax$
(see Swaters  \etal 2003;  Rhee \etal 2004;  Hayashi \&  Navarro 2006;
Valenzuela \etal 2007), which will  cause the opposite effect and thus
would favor cosmologies with higher $\sigma_8$.

For the  WMAP5 cosmology  we show the  1 and $2-\sigma$  dispersion in
$\Delta_{V/2}$   assuming  a   scatter  in   the  $c-M$   relation  of
$\sigma_{\ln c}=0.26$ (which results in a scatter of $\simeq 0.28$ dex
in $\Delta_{V/2}$).  The dispersion  of the observed $\Delta_{V/2}$ is
$\simeq 0.43$  dex. However, subtracting a  measurement uncertainty of
0.32 dex  (the median  measurement uncertainty from  the observations)
results  in  an  intrinsic  scatter  of $\simeq  0.29$  dex,  in  good
agreement with our theoretical predictions.

We conclude therefore, that modulo the caveats of halo contraction and
systematic effects (which  to first order tend to  cancel each other),
the central densities of dwarf  and LSB galaxies are in good agreement
with predictions for a \LCDM  cosmology with parameters favored by the
WMAP mission.

However,  on  larger  scales   observations  seem  to  require  higher
concentrations than  predicted by  \LCDM.  Using X-ray  ({\it Chandra}
and  {\it XMM-Newton})  observations  of 39  massive galaxy  clusters,
Buote    \etal    (2007)    obtained    $\cvir=9.0   \pm    0.4$    at
$\Mvir=10^{14}\Msun$ for  their full  sample (covering the  mass range
$10^{13} \lta  \Mvir \lta 10^{15} \Msun$),  and $\cvir =  7.6 \pm 0.5$
for the  sample restricted to  $\Mvir \geq 10^{14}  \Msun$.  Combining
strong lensing  and X-ray observations, Comerford  \& Natarajan (2007)
found $\cvir=11.1$  at $\Mvir=10^{14}  \Msun$.  In our  simulations we
find    the   median   concentrations    of   relaxed    haloes   with
$\Mvir=10^{14}\Msun$  to  be  $\cvir=6.9$  (WMAP1)  and  $\cvir=  5.5$
(WMAP3).

The median $\cvir$ in the full  sample of Buote \etal is a factor 1.30
higher than in our WMAP1 simulations  and a factor 1.63 higher than in
our WMAP3  simulations.  For the high  mass sample of  Buote \etal the
discrepancies  are smaller: a  factor of  1.1 for  WMAP1 and  1.25 for
WMAP3. Given  that the scatter in  halo concentrations is  a factor of
1.3, selection biases in the  data (such as selecting the most relaxed
clusters, which  are likely  to form earlier  and thus to  have higher
concentration) could  plausibly reconcile the  observed concentrations
from  Buote \etal  with the  WMAP3 cosmology.   The  concentrations of
Comerford \& Natarajan  (2007) are a factor 1.6  higher than our WMAP1
results  and a  factor 2.0  higher than  WMAP3 results,  and selection
effects would  have to  be rather severe  to reconcile even  our WMAP1
results  with these  observations.  Another  possibility is  that halo
contraction due  to the condensation of  baryons at the  center of the
halo (Blumenthal  \etal 1986) has played an  important role.  However,
studies  which  model the  radial  density  profiles  of clusters  and
elliptical galaxies  suggest that it is difficult  to reconcile models
with adiabatic  contraction with observations  (Zappacosta \etal 2006;
Humphrey  \etal 2006, Gastaldello  \etal 2007).   Thus, it  seems that
there is  a discrepancy between model  and data at the  high mass end,
but not at the low mass end.

This may  signal the  need for different  normalizations of  the power
spectrum  on different  scales, which  in turn  may indicate  that the
power spectrum  has a  significant tilt, or  a running  spectral index
with  a  shallower  slope  (lower  $n$)  at  lower  masses.   However,
systematic  errors  and selection  effects  in  the  data need  to  be
understood better before such a conclusion can be verified.

\begin{center}
\begin{figure}
\psfig{figure=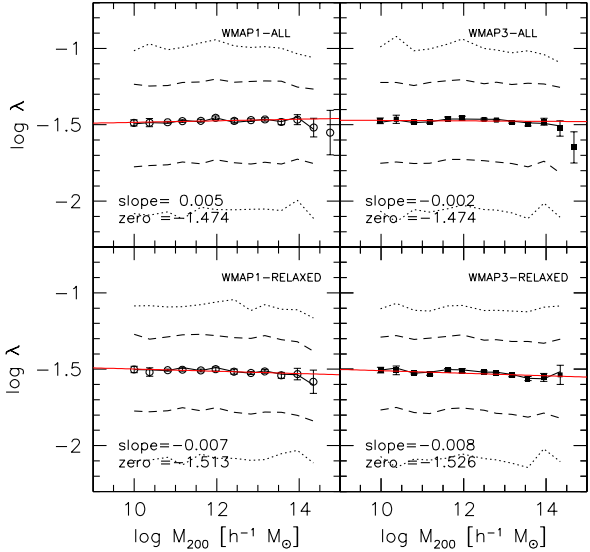,width=0.47\textwidth}
\caption{\scriptsize  Spin parameter  vs.  mass for  all halos  (upper
  panels) and relaxed halos (lower panels). WMAP5 result are not shown
  since they do  not present any significant difference  from WMAP1 or
  WMAP3.  The points  represent the median spin in  each mass bin, the
  error  bar shows  the Poisson  error on  the mean.   The  dashed and
  dotted lines show the 15.9,  84.1, 2.3 and 97.7th percentiles of the
  distribution.  The  solid (red)  line shows a  power-law fit  to the
  $\lambda- M_{200}$ relation: $\log \lambda = {\rm zero + slope}(\log
  M_{200}/12 h^{-1}M_{\odot})$ whose parameters are given in the lower
  left  corner of  each panel.   Linear fit  parameters for  the WMAP5
  model are reported in Table \ref{tab:fits1}.  }
\label{fig:lm4s}
\end{figure}
\end{center}

\begin{center}
\begin{figure}
\psfig{figure=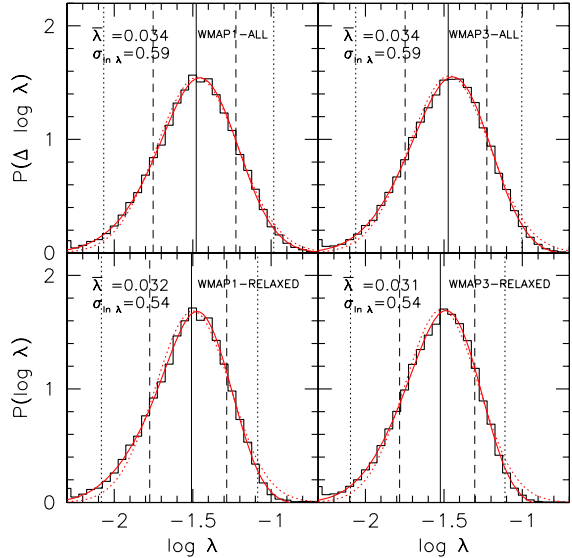,width=0.47\textwidth}
\caption{\scriptsize Histograms  of the distribution of  the halo spin
  parameters.  Only results for WMAP1 and WMAP3 are shown, results for
  WMAP5 are listed in  Table \ref{tab:fits2}.  The vertical lines show
  the 2.3th, 13.9th, 50th, 84.1th,  and 97.7th percentiles of the spin
  residuals.  The  solid red  line  shows  a Gauss-Hermite  polynomial
  expansion up to  fourth order (whose parameters are  listed in Table
  \ref{tab:fits2}),   the   dashed  red   line   shows  the   Gaussian
  corresponding to  the zeroth order  of this expansion. The  mean and
  the  logarithmic  variance  ($\sigma_{\rm  ln \  \lambda}$)  of  the
  Gaussian fit are reported in the top left corner of each panel. }
\label{fig:lm4hist}
\end{figure}
\end{center}

\section{Spin parameters}
\label{sec:spin}

Fig.~\ref{fig:lm4s} shows the spin  parameter versus halo mass for the
WMAP1 (left)  and WMAP3 (right)  cosmologies for all haloes  (top) and
for the  subsample of relaxed haloes  (bottom). As in M07,  we find no
mass or  cosmology dependence of  the halo spin parameters.   For this
reason we  do not include the  results for the WMAP5  cosmology in the
plots.  For  completeness, we do  list the corresponding  parameters in
Tables \ref{tab:fits1} and \ref{tab:fits2} in the Appendix.

Fig.~\ref{fig:lm4hist} shows  the distribution of  spin parameters. As
with the  scatter in  concentration, the scatter  in $\lambda$  is not
perfectly log-normal.  The distribution  of spins for all haloes shows
a small skewness to low  spin parameters.  However, we caution against
an over interpretation of this skewness, as Bullock \etal (2001b) have
demonstrated  that  haloes with  lower  spins  have larger  associated
uncertainties,  with the  error  in $\lambda$  given  roughly by  $0.2
\lambda^{-1} N^{-1/2}$.  Thus a halo  with 1000 particles will have an
uncertainty  of 63\%,  18\%, and  6\% for  a spin  parameter  of 0.01,
0.035, and 0.10.  Even for haloes  with 10000 particles the error on a
spin parameter of 0.01 is  20\%, which will introduce a non-negligible
skewness to the distribution of $\lambda$.

We find  that the mean,  dispersion and skewness of  the distributions
(see  Table  \ref{tab:fits2}) are  remarkably  similar  for all  three
cosmologies considered here,  both for the full set  of haloes and for
the subsample  of relaxed haloes.   The relaxed haloes have  a smaller
mean and variance, and a larger  skewness.  This is due to the removal
of un-relaxed  haloes which typically  have higher spins  than relaxed
haloes (see M07 for details).

\begin{center}
\begin{figure}
\psfig{figure=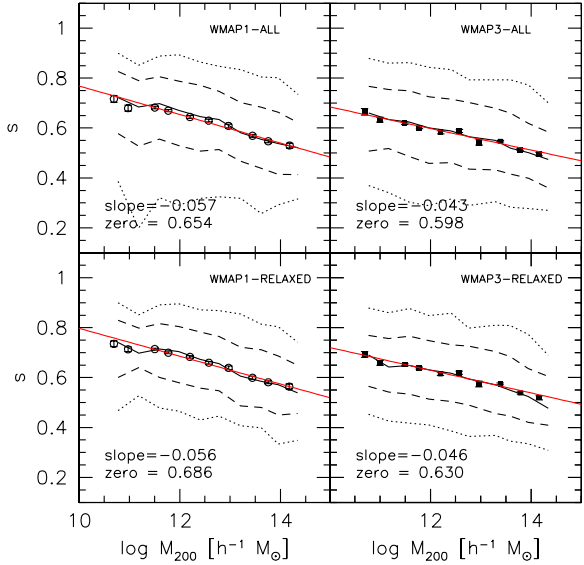,width=0.47\textwidth}
\caption{\scriptsize Short to long axis  ratio, $s$, vs mass for WMAP1
  (left) and  WMAP3 (right),  and for all  haloes (upper)  and relaxed
  haloes  (lower) with  $N_{200} >  3000$.  WMAP5  model  results (not
  shown) lie  in between  WMAP1 and WMAP3.   The points  represent the
  median value of  $s$ spin in each mass bin, the  error bar shows the
  Poisson error  on the  mean.  The dashed  and dotted lines  show the
  15.9,  84.1, 2.3 and  97.7th percentiles  of the  distribution.  The
  solid (red) line shows a  power-law fit to the $s-M_{200}$ relation:
  $s  = {\rm  zero  + slope}(\log  M_{200}/12 h^{-1}M_{\odot})$  whose
  parameters are given in the lower left corner of each panel.  Linear
  fit  parameters   for  the  WMAP5   model  are  reported   in  Table
  \ref{tab:fits1}.}
\label{fig:sm4s}
\end{figure}
\end{center}

\begin{center}
\begin{figure}
\psfig{figure=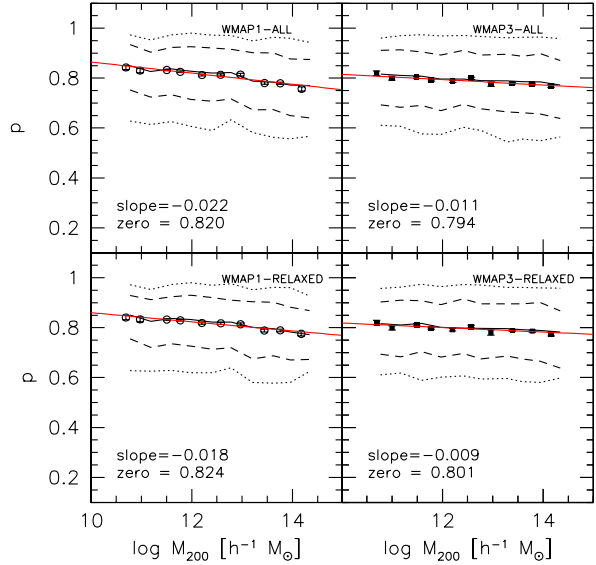,width=0.47\textwidth}
\caption{\scriptsize Same  of Fig.  \ref{fig:sm4s} but for  the middle
  axis, $p$, vs mass.  WMAP5  model results (not shown) lie in between
  WMAP1 and WMAP3.   Values for the {\it slope} and  the {\it zero} of
  the  linear  fit   for  the  WMAP5  model  are   reported  in  Table
  \ref{tab:fits1}.}
\label{fig:pm4s}
\end{figure}
\end{center}

\begin{center}
\begin{figure}
\psfig{figure=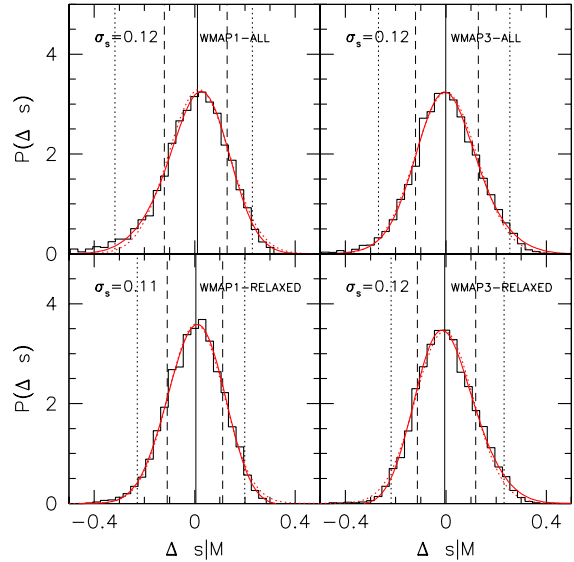,width=0.47\textwidth}
\caption{\scriptsize Histograms of scatter in the halo shape parameter
  $s$ (ratio between minor and major axis length) for WMAP1 (left) and
  WMAP3 (right); results for WMAP5  (not shown in the plot) are listed
  in  Table \ref{tab:fits2}.  The  upper panels  show all  haloes with
  $N_{200} >  3000$, while the  lower panels show the  relaxed haloes.
  The dashed  (red) lines  show the best  fitting Gaussian,  while the
  solid (red) lines show the Gauss-Hermite polynomial expansion (whose
  parameter are  reported in Table \ref{tab:fits2}).   The variance of
  the  Gaussian fit  is  reported in  the  upper left  corner of  each
  panel.}
\label{fig:sm4hist}
\end{figure}
\end{center}

\begin{center}
\begin{figure}
\psfig{figure=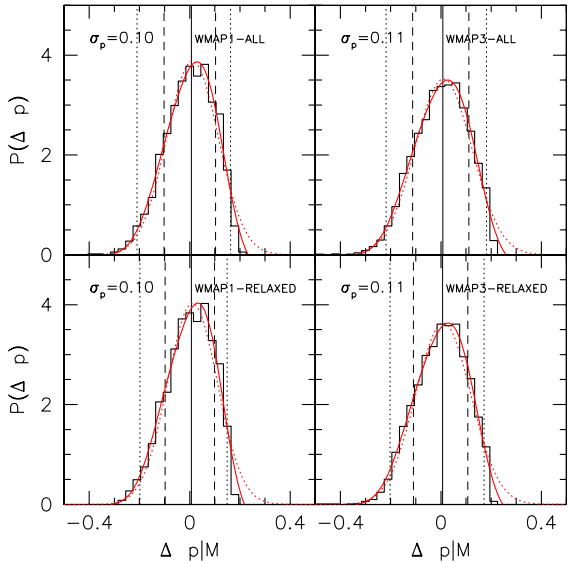,width=0.47\textwidth}
\caption{\scriptsize Histograms of scatter in the halo shape parameter
  $p$  (ratio between minor  and intermediate  axis length)  for WMAP1
  (left) and WMAP3 (right); results  for WMAP5 (not shown in the plot)
  are  listed in  Table \ref{tab:fits2}.   The upper  panels  show all
  haloes with greater than  3000 particles within $r_{200}$, while the
  lower panels  show the ``relaxed''  haloes.  The dashed  (red) lines
  show the best fitting Gaussian, while the solid (red) lines show the
  Gauss-Hermite  polynomial  expansion   (parameters  listed  in  Table
  \ref{tab:fits2}).  The  variance of the Gaussian fit  is reported in
  the upper left corner of each panel.}
\label{fig:pm4hist}
\end{figure}
\end{center}

\section{Halo Shapes}
\label{sec:shape}

Fig  ~\ref{fig:sm4s} shows the  relation between  $s$ (defined  as the
ratio between  the short and long  axes) and $M_{200}$.   Here we only
consider  haloes with  at  least 3000  particles,  because with  fewer
particles we  find evidence  for resolution effects.   The $s-M_{200}$
relation is well  fitted with a power-law of  slope $\simeq -0.05$, in
all cosmologies  and for all  and relaxed haloes.   Fitting parameters
for  the  shape-mass relation  and  for  its  scatter, for  all  three
cosmological  models,   are  listed  in   Tables  \ref{tab:fits1}  and
\ref{tab:fits2} in  the Appendix.  For  the WMAP3 cosmology,  the zero
point for  relaxed haloes is 0.03  higher than for the  full sample of
haloes. In  addition, we find that  the zero-point if  0.05 higher for
the WMAP1 cosmology  compared to that of the  WMAP1 cosmology.  On the
other hand, the relation between $p$ (defined as the ratio between the
short  and  intermediate  axes)  and  $M_{200}$ shows  a  much  weaker
correlation,   and   only   a   marginal   dependence   on   cosmology
(Fig.~\ref{fig:pm4s}).

Figs.~\ref{fig:sm4hist}  \& ~\ref{fig:pm4hist} show  the distributions
of $s$ and $p$ about  the mean relations shown in Figs.~\ref{fig:sm4s}
\&  ~\ref{fig:pm4s}.  These  are  roughly Gaussian.   The full  sample
reveals a  mild skewness to low $s$,  which is most likely  due to the
presence of  unrelaxed haloes.  

Allgood \etal  (2006) presented a detailed analysis  of the dependence
of halo shape  on mass and on the  underlying cosmological model. They
found that  the redshift, mass  and $\sigma_8$ dependence of  the mean
smallest-to-largest axis ratio of haloes is well described by a simple
power  law relation  $\langle  s_{0.3} \rangle  = a  (\Mvir/M_{*})^b$,
where in  this case $s_{0.3}$  is measured inside $0.3\Rvir$,  and the
$z$  and $\sigma_8$  dependencies are  governed by  the characteristic
non-linear  mass,  $M_*=M_*(z,\sigma_8)$.   Using several  simulations
they found  the following values  for the fitting  parameters: $a=0.54
\pm 0.02$ and $b=-0.050 \pm 0.003$.

\begin{center}
\begin{figure}
\psfig{figure=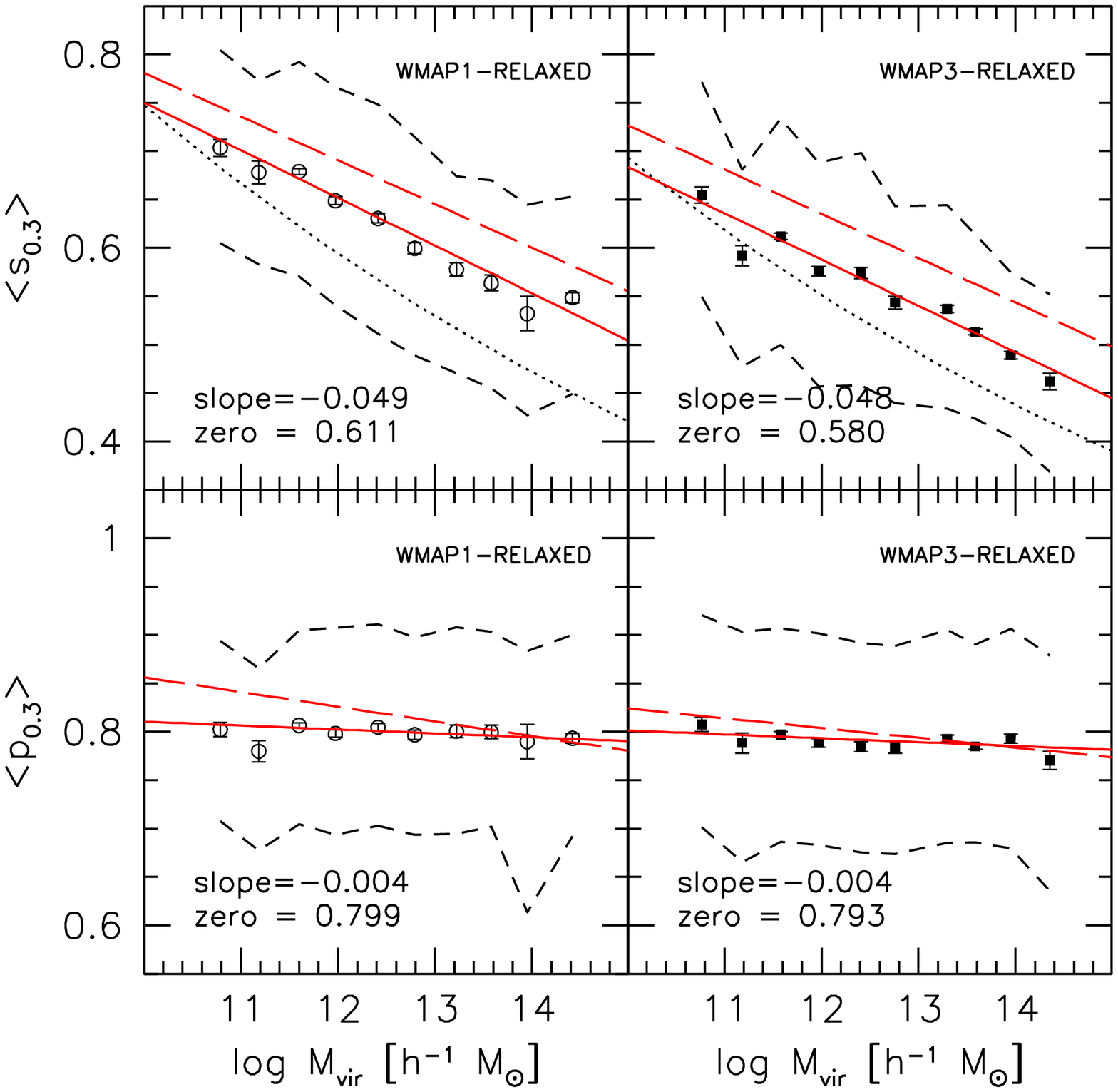,width=0.47\textwidth}
\caption{\scriptsize Shape within  0.3$\Rvir$ versus mass in different
  cosmological   models.   A  universe   with  lower   $\sigma_8$  and
  $\Omega_m$  produces haloes that  are more  elongated. In  the upper
  panels the  dotted lines give the relationship  suggested by Allgood
  \etal (2006).  The solid lines show  a fit of the form shape $= {\rm
    zero  + slope}(\log  \Mvir/M_*)$,  where $\log  M_*= 12.82,  12.17
  \hMsun$  in  the WMAP1  and  WMAP3  cosmologies, respectively.   The
  slopes and  zero points of  these fits are  given in the  lower left
  corner of each panel.   The long-dashed lines show the corresponding
  relations  for the shapes  measured within  the virial  radius.  The
  short dashed lines enclose 68.3\% of the points, and correspond to a
  scatter of $\simeq 0.1$.}
\label{fig:s3mvir}
\end{figure}
\end{center}

Fig.~\ref{fig:s3mvir} shows the mean $s_{0.3}$-mass and $p_{0.3}$-mass
relations in our simulations. The dotted line shows the relation found
by Allgood  \etal (2006)  where we used  the following values  for the
characteristic  mass: $\rm  log(M_{*}/  h^{-1} \rm  M_{\odot})=$12.82,
12.17,  for  the  WMAP1  and  WMAP3  cosmologies,  respectively.   Our
simulations show  a similar  slope, but somewhat  higher normalization
compared  to Allgood  \etal (2006).   In  addition, we  find that  the
differences in the shapes between  WMAP1 and WMAP3 cosmologies are not
simply  explained by  a difference  in $M_*$,  as proposed  by Allgood
\etal   (2006).   This   can  be   seen   in  the   upper  panels   of
Fig.~\ref{fig:s3mvir} by comparing  the difference between the Allgood
\etal  (2006)  prediction  (dashed  line)  with  the  results  of  our
simulations  (solid  line).   This  difference  is  not  constant  and
increases from WMAP1 to WMAP3.

The dot-dashed lines  in Fig \ref{fig:s3mvir} show $s$  and $p$ within
the virial radius.   This shows that $s$ decreases  towards the center
of  the  halo, in  good  agreement  with  Allgood \etal  (2006).   The
dispersion in  the axis  ratios measured at  0.3$\Rvir$ is  0.10, only
marginally  smaller  than  the  dispersions measured  within  $\Rvir$.
Results for the shape-mass relation and for the shape distribution for
the WMAP5  model (not  shown in the  figures) are summarized  in Table
\ref{tab:fits1} and in Table \ref{tab:fits2}.

\section{Summary}
\label{sec:conc}

In  this  paper  we have  used  a  large  set of  cosmological  N-body
simulations in WMAP1, WMAP3 and WMAP5 cosmologies to study how changes
in the cosmological parameters affect  the halo mass dependence of the
concentration parameter, $c$, the spin parameter, $\lam$, and the halo
shape  parameters, $s$  and $p$.   The  simulations span  5 orders  of
magnitude in halo mass ($10^{10}-10^{15} \hMsun$), covering the entire
range from haloes  those that host individual dwarf  galaxies to those
associated with massive clusters.

At a  fixed mass, haloes in  a WMAP3 cosmology  are significantly less
concentrated than their counterparts in a WMAP1 cosmology.  As already
noted by  other authors this  can be ascribed  to the lower  value for
$\sigma_8$ in the  WMAP3 cosmology compared to WMAP1,  that shifts the
entire  process of  structure formation  towards lower  redshifts.  In
particular, for  a halo of  $10^{12} \hMsun$ the  WMAP3 concentrations
are $1.41$  times lower than in  a WMAP1 cosmology. Due  to this lower
normalization,  the central  densities of  dark matter  haloes  in the
WMAP3 cosmology are consistent  with the observed central densities of
dark matter haloes  of dwarf and LSB galaxies.   However, on the scale
of clusters, the WMAP3 concentrations may actually be too low compared
to  observational   constraints  from  X-ray   measurements  and  from
gravitational  lensing.   The  WMAP5 concentrations  are  intermediate
between  the WMAP1 and  WMAP3 cosmologies:  they are  still consistent
with the data  on dwarf and LSB galaxies, but somewhat  too low on the
scale of galaxy  clusters.  Although this may indicate  a problem with
the  exact  shape of  the  power  spectrum  of density  perturbations,
systematic errors and selection effects in the cluster data need to be
better  understood, before  the  data  can be  used  to constrain  the
cosmological spectral index.

We find that for all three cosmologies, the average halo concentration
as function of halo mass is well fitted by a single power-law over the
entire range of halo masses  covered by our simulations.  This is {\it
  inconsistent} with  all existing models  for the mass  dependence of
halo  concentrations, which  predict  that the  slope  of the  $c$-$M$
relation becomes  steeper at higher masses.  In  particular, the model
suggested by  Navarro, Frenk, \&  White (1997) matches the  slopes and
(to a  lesser degree)  normalizations of the  $c$-$M$ relation  at the
massive end ($M \gta  10^{13} \hMsun$), but dramatically underpredicts
the  concentrations  for low  mass  haloes.   The  model suggested  by
Bullock \etal  (2001a), on the other  hand, matches the  slopes of the
$c$-$M$ relations  for $M \lta 10^{13} \hMsun$,  but underpredicts the
concentrations for more massive haloes.

We propose a modification to the B01 model, based on the idea that the
characteristic  density of  a  halo remains  constant  after the  halo
forms.  This results in a growth in halo concentration proportional to
$H(z_c)^{2/3}$,   rather  than   $(1+z_c)$  as   in  the   B01  model.
Consequently, our new model reproduces the slope of the $c-M$ relation
over  the  full range  of  masses in  our  simulations,  and for  each
cosmology. However,  as for the original B01  model, the normalization
is cosmology  dependent (at  the level of  a few  percent), suggesting
that  the  model  still  does  not  completely  capture  the  relevant
dependencies on the power spectrum.   Until such a model is available,
the normalization of the $c$-$M$  relations will have to be calibrated
against  numerical  simulations   for  each  cosmology,  if  precision
concentration parameters are required.

In agreement with previous studies  (e.g. Bullock \etal 2001b) we find
that the distribution  of spin parameters is independent  of halo mass
and cosmology. There is a trend, though, that less relaxed haloes have
higher  spin  parameters,  as  previously noticed  by  Macci\`o  \etal
(2007).  Finally, we find  haloes to  be more  flattened in  the WMAP3
cosmology than in the  WMAP1 cosmology, consistent with the suggestion
that haloes that form later are more aspherical (Allgood \etal 2006).

\section*{Acknowledgements} 

We thank  Andrey Kravtsov for  helpful discussions, and  James Bullock
for  providing  a  publically  available  version of  his  code  which
includes            our                       modifiication
(http://www.physics.uci.edu/\~{}bullock/CVIR/).
The numerical simulations were performed on the zBox2 supercomputer at
the  University   of  Z\"urich   and  on  the   PIA  cluster   of  the
Max-Planck-Institut f\"ur Astronomie at the Rechenzentrum in Garching.
Special thanks to Ben Moore,  Doug Potter  and Joachim  Stadel for  
bringing zBox2  to life.
A.A.D.   acknowledges  support from  the  National Science  Foundation
Grant AST-0507483.


\appendix
\section{parameters}

In  the  appendix we  summarize  all  the  parameters of  the  fitting
functions used through this  paper. Table \ref{tab:fits1} contains the
slope, zero and relative errors for the power-laws fits shown in Figs.
\ref{fig:cm4s_vir},         \ref{fig:cm4s_200},        \ref{fig:lm4s},
\ref{fig:sm4s}.  Table \ref{tab:fits2} shows the parameters describing
the distribution around the mean of concentrations, spin and shapes in
the  three  different  cosmological models  (Figs:  \ref{fig:cm4hist},
\ref{fig:lm4hist}, \ref{fig:sm4hist},  \ref{fig:pm4hist}).  While this
paper was  ready for submission  a similar study of  the concentration
mass-relation  for the WMAP5  cosmology was  presented in  Duffy \etal
(2008). They  limited their studied  only to haloes defined  with more
than $10^4$ particles  inside the virial radius.  This  left them with
only 1269  haloes, nevertheless  the slope of  their power law  fit is
consistent with our  results (within the errors) for  both the ALL and
RELAXED sample.

\begin{table*}
  \caption{Parameters of the power-law fits shown in Figs.\ref{fig:cm4s_vir}, \ref{fig:cm4s_200}, 
    \ref{fig:lm4s}, \ref{fig:sm4s}, \& \ref{fig:pm4s}. The data are fitted with the power law 
    $y= $\rm zero$ + $\rm slope$ (\log M -12)$, where $M$ is in units of $\hMsun$. 
    The second column reports the mass range over which the fit has been obtained, the third 
    the number of haloes within this mass range (using 
    all the simulations listed in Table \ref{tab:sims}).
  }
\begin{tabular}{lrrrrrrr}
\hline
sample & $N_{min}$ & $\log_{10} M$ & ${\rm N}_{\rm haloes}$ &  zero & error & slope & error \\ 
                  & & $[h^{-1}\, M_\odot]$ \\
\hline
$\log c_{\rm vir}\, vs \, \log M_{\rm vir}$ \\
W1-ALL     &   500 & 9.83-14.95 & 29693 & 1.011 & 0.001 & -0.114 & 0.001 \\
W1-RELAXED &   500 & 9.83-14.95 & 21876 & 1.051 & 0.001 & -0.099 & 0.001 \\
W3-ALL     &   500 & 9.82-14.95 & 49830 & 0.861 & 0.001 & -0.086 & 0.001 \\
W3-RELAXED &   500 & 9.82-14.95 & 33913 & 0.915 & 0.001 & -0.080 & 0.001 \\
W5-ALL     &   500 & 9.84-14.86 & 12184 & 0.925 & 0.001 & -0.108 & 0.001 \\
W5-RELAXED &   500 & 9.84-14.86 &  8282 & 0.971 & 0.001 & -0.094 & 0.001 \\
\hline
$\log c_{200}\, vs \, \log M_{200}$ \\
W1-ALL     &   500 & 9.83-14.92 & 25952 & 0.879 & 0.001 & -0.119 & 0.001 \\
W1-RELAXED &   500 & 9.83-14.85 & 19528 & 0.917 & 0.001 & -0.104 & 0.001 \\
W3-ALL     &   500 & 9.82-14.82 & 40027 & 0.719 & 0.001 & -0.088 & 0.001 \\
W3-RELAXED &   500 & 9.82-14.82 & 28344 & 0.769 & 0.001 & -0.083 & 0.001 \\
W5-ALL     &   500 & 9.84-14.93 &  9988 & 0.787 & 0.001 & -0.110 & 0.001 \\
W5-RELAXED &   500 & 9.84-14.93 &  7060 & 0.830 & 0.001 & -0.098 & 0.001 \\
\hline
$\log \lambda\, vs \, \log M_{200}$ \\
W1-ALL     &   500 & 9.83-14.92 & 25733 &-1.474 & 0.002 & 0.005 & 0.002\\
W1-RELAXED &   500 & 9.83-14.92 & 19377 &-1.513 & 0.002 &-0.007 & 0.002\\
W3-ALL     &   500 & 9.82-14.82 & 39662 &-1.474 & 0.001 &-0.002 & 0.001\\
W3-RELAXED &   500 & 9.82-14.82 & 28114 &-1.526 & 0.001 &-0.008 & 0.001\\
W5-ALL     &   500 & 9.84-14.93 &  9988 &-1.458 & 0.004 & 0.001 & 0.003 \\
W5-RELAXED &   500 & 9.84-14.93 &  7060 &-1.505 & 0.004 &-0.009 & 0.004 \\
\hline
$s \, vs \, \log M_{200}$ \\
W1-ALL     &  3000 & 10.61-14.92 & 4886 & 0.654 & 0.015 & -0.057 & 0.016\\
W1-RELAXED &  3000 & 10.61-14.85 & 3820 & 0.686 & 0.017 & -0.056 & 0.018\\
W3-ALL     &  3000 & 10.60-14.82 & 6022 & 0.598 & 0.016 & -0.043 & 0.013\\
W3-RELAXED &  3000 & 10.60-14.82 & 4273 & 0.630 & 0.018 & -0.046 & 0.016\\
W5-ALL     &  3000 & 10.60-14.82 & 1492 & 0.623 & 0.035 & -0.052 & 0.025\\
W5-RELAXED &  3000 & 10.61-14.93 & 1060 & 0.657 & 0.040 & -0.054 & 0.029\\
\hline
$p \, vs \, \log M_{200}$ \\
W1-ALL     &  3000 & 10.61-14.92 & 4886 & 0.820 & 0.015 & -0.022 & 0.016\\
W1-RELAXED &  3000 & 10.61-14.85 & 3820 & 0.824 & 0.017 & -0.018 & 0.018\\
W3-ALL     &  3000 & 10.60-14.82 & 6022 & 0.794 & 0.016 & -0.011 & 0.013\\
W3-RELAXED &  3000 & 10.60-14.82 & 4273 & 0.801 & 0.018 & -0.009 & 0.016\\
W5-ALL     &  3000 & 10.60-14.82 & 1492 & 0.807 & 0.035 & -0.016 & 0.025\\
W5-RELAXED &  3000 & 10.61-14.93 & 1060 & 0.812 & 0.040 & -0.014 & 0.029\\
\hline
\end{tabular}
\label{tab:fits1}
\end{table*}

\begin{table*}
  \caption{Parameters of the distribution of concentrations (Fig \ref{fig:cm4hist}) , spin (Fig.~\ref{fig:lm4hist}), and shapes (Figs.~\ref{fig:sm4hist} \& \ref{fig:pm4hist}).
    The distribution have been fitted by a Gauss-Hermite polynomial expansion up to fourth order 
    ($h_1-h_4$). The zeroth order of this expansion is a Gaussian fit, whose mean $<\Delta\log c>$
     and dispersion $\sigma_{\Delta\log c}$ are determined by setting $h_1=h_2=0$.
    Columns 6-10 show the 2.3, 15.9, 50.0, 84.1, and  97.7th percentiles  of the  distribution.
  }

\begin{tabular}{lrrrrrrrrrrrrr}
\hline
sample & $< \Delta \log c> $        & $\sigma_{\Delta \log c}$            & $h_3$ & $h_4$ & 2.3th & 15.9th & 50th & 84.1th & 97.7th\\
\hline
$\Delta \log c_{200} | M_{200}$ \\
W1-ALL     &0.033 & 0.129 & -0.670 &  0.356 & -0.405 & -0.143 & 0.022 & 0.146 & 0.269\\
W1-RELAXED &0.011 & 0.111 & -0.421 &  0.216 & -0.263 & -0.113 & 0.007 & 0.115 & 0.228\\
W3-ALL     &0.040 & 0.132 & -0.880 &  0.252 & -0.417 & -0.147 & 0.027 & 0.150 & 0.252\\
W3-RELAXED &0.015 & 0.109 & -0.622 &  0.146 & -0.260 & -0.113 & 0.010 & 0.114 & 0.209\\
W5-ALL     &0.041 & 0.130 & -0.929 &  0.351 & -0.417 & -0.148 & 0.028 & 0.149 & 0.252\\
W5-RELAXED &0.015 & 0.105 & -0.620 &  0.259 & -0.268 & -0.110 & 0.011 & 0.112 & 0.209\\
\hline
$\log \lambda$ \\
W1-ALL     &-1.470& 0.258 & -0.158 & 0.050 & -2.069 & -1.752 & -1.478 & -1.225 & -0.984\\
W1-RELAXED &-1.499& 0.236 & -0.307 & 0.034 & -2.083 & -1.774 & -1.508 & -1.284 & -1.085\\
W3-ALL     &-1.465& 0.256 & -0.194 &-0.003 & -2.062 & -1.747 & -1.472 & -1.227 & -1.004\\
W3-RELAXED &-1.514& 0.235 & -0.297 &-0.019 & -2.095 & -1.783 & -1.522 & -1.303 & -1.110\\
W5-ALL     &-1.466& 0.253 & -0.162 & 0.051 & -2.070 & -1.743 & -1.468 & -1.218 & -0.942\\
W5-RELAXED &-1.508& 0.228 & -0.341 & 0.041 & -2.105 & -1.774 & -1.515 & -1.297 & -1.065\\
\hline
$\Delta s | M_{200}$ \\
W1-ALL     & 0.017 & 0.121 &-0.343 & 0.098 & -0.317 & -0.121 & 0.011 & 0.130 & 0.230\\
W1-RELAXED & 0.004 & 0.112 &-0.123 &-0.123 & -0.224 & -0.110 & 0.003 & 0.112 & 0.205\\
W3-ALL     & 0.007 & 0.123 & 0.215 & 0.169 & -0.269 & -0.121 &-0.001 & 0.129 & 0.255\\
W3-RELAXED &-0.006 & 0.116 & 0.441 & 0.011 & -0.218 & -0.113 &-0.004 & 0.119 & 0.232\\
W5-ALL     & 0.008 & 0.128 &-0.117 & 0.196 & -0.294 & -0.129 & 0.004 & 0.133 & 0.248\\
W5-RELAXED &-0.002 & 0.117 & 0.321 &-0.031 & -0.233 & -0.114 &-0.003 & 0.123 & 0.221\\
\hline
$\Delta p | M_{200}$ \\
W1-ALL     &0.011 & 0.103 & -0.779 &-0.514 & -0.209 & -0.102 & 0.006 & 0.103 & 0.162\\
W1-RELAXED &0.011 & 0.100 & -0.867 &-0.504 & -0.203 & -0.099 & 0.005 & 0.100 & 0.155\\
W3-ALL     &0.010 & 0.113 & -0.596 &-0.491 & -0.218 & -0.113 & 0.008 & 0.112 & 0.181\\
W3-RELAXED &0.010 & 0.110 & -0.731 &-0.525 & -0.203 & -0.109 & 0.007 & 0.108 & 0.173\\
W5-ALL     &0.009 & 0.109 & -0.477 &-0.335 & -0.218 & -0.109 & 0.004 & 0.106 & 0.185\\
W5-RELAXED &0.006 & 0.103 & -0.451 &-0.287 & -0.198 & -0.101 & 0.005 & 0.100 & 0.178\\

\hline
\end{tabular}
\label{tab:fits2}
\end{table*}

\label{lastpage}


\begin{thebibliography}{}

\bibitem[Alam   et  al.(2002)]{2002ApJ...572...34A}   Alam  S.~M.~K.,
  Bullock J.~S., Weinberg D.~H.\ 2002, \apj, 572, 34

\bibitem[Allgood et al.(2006)]{2006MNRAS.367.1781A} Allgood B., Flores 
R.~A., Primack J.~R., Kravtsov A.~V., Wechsler R.~H., Faltenbacher A., 
 Bullock J.~S.\ 2006, MNRAS, 367, 1781 



\bibitem[Bertschinger(2001)]{2001ApJS..137.,1B} Bertschinger E.\ 2001, 
ApJS, 137, 1 

\bibitem[Blumenthal   et  al.(1986)]{1986ApJ...301...27B}  Blumenthal
G.~R., Faber S.~M.,  Flores R., Primack J.~R.\  1986, \apj, 301,
27

\bibitem[Bullock  et  al.(2001a)]{2001MNRAS.321..559B} Bullock  J.~S.,
Kolatt T.~S., Sigad Y.,  Somerville R.~S., Kravtsov A.~V., Klypin
A.~A., Primack J.~R.,  Dekel A.\ 2001a, MNRAS, 321, 559 (B01)

\bibitem[Bullock  et  al.(2001b)]{2001ApJ...555..240B} Bullock  J.~S.,
Dekel A.,  Kolatt T.~S.,  Kravtsov A.~V., Klypin  A.~A., Porciani
C.,  Primack J.~R.\ 2001b, ApJ, 555, 240

\bibitem[Buote et al.(2007)]{2007ApJ...664..123B} Buote D.~A., 
Gastaldello F., Humphrey P.~J., Zappacosta L., Bullock J.~S., 
Brighenti F.,  Mathews W.~G.\ 2007, \apj, 664, 123 


\bibitem[Comerford  Natarajan(2007)]{2007MNRAS.379..190C} Comerford J.~M.,  Natarajan P.\ 2007, \mnras, 379, 190 

\bibitem[de Blok et al.(2001)]{2001AJ.,122.2396D} de Blok W.~J.~G., 
McGaugh S.~S.,  Rubin V.~C.\ 2001, AJ, 122, 2396 

\bibitem[de   Blok    Bosma(2002)]{2002AA...385..816D}   de  Blok
  W.~J.~G.,  Bosma A.\ 2002, A\&A, 385, 816



\bibitem[Duffy (2008)]{2005ApJ...619..218D}  Duffy A., Schaye J., Kay S., 
Dalla Vecchia C. 2008, arXiv:0804:2486, MNRAS in press


\bibitem[Dutton  et   al.(2005)]{2005ApJ...619..218D}  Dutton  A.~A.,
Courteau S., de Jong R.,  Carignan C.\ 2005, \apj, 619, 218

\bibitem[Dutton et al.(2007)]{2007ApJ...654...27D} Dutton A.~A., van den 
Bosch F.~C., Dekel A.,  Courteau S.\ 2007, \apj, 654, 27 

\bibitem[Eke  et al.(2001)]{2001ApJ...554..114E} Eke  V.~R., Navarro
J.~F.,  Steinmetz M.\ 2001, ApJ, 554, 114

\bibitem[Flores  Primack(1994)]{1994ApJ...427L...1F} 
Flores R.~A.,  Primack J.~R.\ 1994, ApJL, 427, L1 

\bibitem[Gastaldello et al.(2007)]{2007ApJ...669..158G} Gastaldello, F., 
Buote, D.~A., Humphrey, P.~J., Zappacosta, L., Bullock, J.~S., Brighenti, 
F., \& Mathews, W.~G.\ 2007, \apj, 669, 158 

\bibitem[Gentile et al.(2005)]{2005ApJ...634L.145G} Gentile, G., Burkert, 
A., Salucci, P., Klein, U., \& Walter, F.\ 2005, ApJL, 634, L145 

\bibitem[Hayashi   Navarro(2006)]{2006MNRAS.373.1117H}  Hayashi E.,
   Navarro J.~F.\ 2006, \mnras, 373, 1117


\bibitem[Humphrey et al.(2006)]{2006ApJ...646..899H} Humphrey P.~J., 
Buote D.~A., Gastaldello F., Zappacosta L., Bullock J.~S., Brighenti 
F.,  Mathews W.~G.\ 2006, \apj, 646, 899 

\bibitem[Jing  Suto(2002)]{2002ApJ...574..538J} Jing Y.~P.,  Suto Y.\ 
2002, \apj, 574, 538


\bibitem[Komatsu et al.(2008)]{2008arXiv0803.0547K} Komatsu E., et al.\ 
2008, ArXiv e-prints, 803, arXiv:0803.0547 

\bibitem[Kuzio de Naray et al.(2008)]{2008ApJ...676..920D} Kuzio de Naray 
R., McGaugh S.~S.,  de Blok W.~J.~G.\ 2008, \apj, 676, 920 

\bibitem[Kuhlen et al.(2005)]{2005MNRAS.357..387K} Kuhlen M., Strigari 
L.~E., Zentner A.~R., Bullock J.~S.,  Primack J.~R.\ 2005, MNRAS, 
357, 387 

\bibitem[Lacey  Cole(1993)]{1993MNRAS.262..627L} Lacey C.,  Cole S.\ 
1993, MNRAS, 262, 627 

\bibitem[Li et al.(2007)]{2007MNRAS.379..689L}Li Y., Mo H.J., 
van den Bosch F.C., Lin W.P., 2007, MNRAS, 379, 689

\bibitem[Macci{\`o} et al.(2003)]{2003ApJ...588...35M} Macci{\`o} A.~V., 
Murante G.,  Bonometto S.~A.\ 2003, ApJ, 588, 35 

\bibitem[Macci{\`o} et al.(2007)]{2007MNRAS.378...55M} Macci{\`o} A.~V., 
Dutton A.~A., van den Bosch F.~C., Moore B., Potter D.,  Stadel J.\ 
2007, MNRAS, 378, 55 (M07)

\bibitem[Mainini et al.(2003)]{2003ApJ...599...24M} Mainini R., 
Macci{\`o} A.~V., Bonometto S.~A.,  Klypin A.\ 2003, ApJ, 599, 24 
 
\bibitem[Mo et al.(1998)]{1998MNRAS.295..319M}  Mo H.~J., Mao S., 
White S.~D.~M.\ 1998, MNRAS, 295, 319

\bibitem[Moore(1994)]{1994Natur.370..629M} Moore B.\ 1994, Nature, 370, 629 


\bibitem[Navarro et al.(1996)]{1996ApJ...462..563N} Navarro J.~F., Frenk 
C.~S.,  White S.~D.~M.\ 1996, \apj, 462, 563 

\bibitem[Navarro  et  al.(1997)]{1997ApJ...490..493N} Navarro  J.~F.,
Frenk C.~S.,  White S.~D.~M.\ 1997, ApJ, 490, 493 (NFW)

\bibitem[Neto et al.(2007)]{2007MNRAS.381.1450N} Neto A.~F., et al.\ 2007, 
\mnras, 381, 1450 

\bibitem[Press \& Schechter(1974)]{1974ApJ...187..425P} Press W.~H.,  
Schechter P.\ 1974, \apj, 187, 425 

\bibitem[Rhee et al.(2004)]{2004ApJ...617.1059R} Rhee G., Valenzuela O., 
Klypin A., Holtzman J.,  Moorthy B., \ 2004, \apj, 617, 1059

\bibitem[Shaw et al.(2006)]{2006ApJ...646..815S} Shaw L.~D., Weller J., 
Ostriker J.~P.,  Bode P.\ 2006, \apj, 646, 815 

\bibitem[Sheth \& Tormen(2002)]{2002MNRAS.329...61S} Sheth R.~K.,  
Tormen G.\ 2002, MNRAS, 329, 61 
 
\bibitem[Simon et al.(2005)]{2005ApJ...621..757S} Simon J.~D., Bolatto 
A.~D., Leroy A., Blitz L.,  Gates E.~L.\ 2005, \apj, 621, 757 

\bibitem[Spergel et al.(2003)]{2003ApJS..148..175S} Spergel D.~N., et
  al.\ 2003, ApJS, 148, 175

\bibitem[Spergel et al.(2007)]{2007ApJS..170..377S} Spergel D.~N., et al.\ 
2007, \apjs, 170, 377 

\bibitem[Springel et al.(2005)]{2005Natur.435..629S} Springel V., et al.\ 
2005, Nature, 435, 629 

\bibitem[Stadel(2001)]{2001PhDT.,.,21S} Stadel J.~G.\ 2001, 
Ph.D.~Thesis, University of Washington  

\bibitem[Swaters et al.(2003)]{2003ApJ...583..732S} Swaters R.~A., Madore 
B.~F., van den Bosch F.~C.,  Balcells M., 2003, ApJ, 583, 732 

\bibitem[Valenzuela et al.(2007)]{2007ApJ...657..773V} Valenzuela O., 
Rhee G., Klypin A., Governato F., Stinson G., Quinn T., 
 Wadsley J.,  2007, \apj, 657, 773 

\bibitem[van den Bosch(2002)]{2002MNRAS.331...98V}
van den Bosch F.C., 2002, MNRAS, 331, 98

\bibitem[van den Bosch \& Swaters(2001)]{2001MNRAS.325.1017V} van den Bosch F.~C.,  Swaters R.~A.\ 2001, \mnras, 325, 1017 

\bibitem[Moore(1994)]{Wec02}
Wechsler R.H., Bullock J.S., Preimack J.R., Kravtsov A.V., Dekel A.,
2002, ApJ, 568, 52 

\bibitem[Zappacosta et al.(2006)]{2006ApJ...650..777Z} Zappacosta L., 
Buote D.~A., Gastaldello F., Humphrey P.~J., Bullock J., Brighenti F., 
 Mathews W.\ 2006, \apj, 650, 777 

\bibitem[Zentner \& Bullock(2002)]{2002PhRvD..66d3003Z} Zentner A.~R.,  
Bullock J.~S.\ 2002, \prd, 66, 043003 

\bibitem[Zhao etal(2003a)]{2003MNRAS...339..12} Zhao D.~H., Mo H.~J.,
Jing Y.~P., B\"orner, G.\ 2003a, MNRAS, 339, 12

\bibitem[Moore(1994)]{Zha03b}
Zhao D.H., Jing Y.P., Mo H.J., B\"orner G., 2003b, ApJ, 597, 9

\end{thebibliography}
\end{document}